\documentclass{article}
\pdfoutput=1

\usepackage{pdfpages}
\usepackage{arxiv}
\usepackage{natbib}
\usepackage{latexsym}
\usepackage{amsmath}
\usepackage{amsfonts}
\usepackage{amssymb}
\usepackage{psfrag}
\usepackage{graphicx}
\usepackage{url}
\pdfoutput=1

\usepackage{mathtools}
\usepackage{bbm}
\usepackage{enumitem}
\usepackage{dsfont}
\usepackage{appendix}
\usepackage{multirow}
\usepackage{subcaption}

\usepackage{graphicx}
\DeclareGraphicsExtensions{.pdf,.png,.jpg}
\graphicspath{ {fig/} }
\pdfoutput=1

\def\Z{\tilde{Y}^*}
\def\Zone{\tilde{A}^*}
\def\Ztwo{\tilde{W}^*}
\def\z{\tilde{y}^*}
\def\zone{\tilde{a}^*}
\def\ztwo{\tilde{w}^*}

\def\bbR{\mathbb{R}}

\def\calL{\mathcal{L}}

\DeclarePairedDelimiterX\bigCond[2]{[}{]}{#1 \;\delimsize\vert\; #2}

\def\part{\pi} 

\title{Modelling multi-scale state-switching functional data with hidden Markov models}

\author{
  Evan Sidrow \\
  Department of Statistics\\
  University of British Columbia\\
  Vancouver, Canada \\
  \texttt{evan.sidrow@stat.ubc.ca} \\
   \And
  Nancy Heckman \\
  Department of Statistics\\
  University of British Columbia\\
  Vancouver, Canada \\
   \And
  Sarah M.E. Fortune \\
  Marine Mammal Research Unit\\
  University of British Columbia\\
  Vancouver, Canada \\
   \And
  Andrew W. Trites \\
  Institute for the Oceans and Fisheries\\
  University of British Columbia\\
  Vancouver, Canada \\
   \And
  Ian Murphy \\
  Department of Statistics\\
  University of British Columbia\\
  Vancouver, Canada \\
   \And
  Marie Auger-M\'eth\'e \\
  Department of Statistics\\
  University of British Columbia\\
  Vancouver, Canada \\
}

\begin{document}
\maketitle

\begin{abstract}
Data sets comprised of sequences of curves sampled at high frequencies in time are increasingly common in practice, but they can exhibit complicated dependence structures that cannot be modelled using common methods of Functional Data Analysis (FDA). We detail a hierarchical approach which treats the curves as observations from a hidden Markov model (HMM). The distribution of each curve is then defined by another fine-scale model which may involve auto-regression and require data transformations using moving-window summary statistics or Fourier analysis. This approach is broadly applicable to sequences of curves exhibiting intricate dependence structures. As a case study, we use this framework to model the fine-scale kinematic movement of a northern resident killer whale ({\em{Orcinus orca}}) off the coast of British Columbia, Canada. Through simulations, we show that our model produces more interpretable state estimation and more accurate parameter estimates compared to existing methods.
\end{abstract}


\section{INTRODUCTION}
\pdfoutput=1

Biologging technology now provides researchers with kinematic data collected almost continuously in time \citep{Hooten:2017}.
The collection and analysis of data from devices such as accelerometers have brought new insights to areas ranging from monitoring machine health \citep{Getman:2009} to understanding physical activity levels in children \citep{Morris:2007}. The study of animal movement in particular has been transformed by tracking devices that record kinematic information in a variety of environments \citep{Borger:2020,Dot:2016b}. Tags can record over 50 observations per second, resulting in time series containing millions of observations over the course of several hours. These data contain a wealth of information about human and animal behaviour, but parsing these large data sets poses a challenge for statisticians and biologists.

Biologging data are frequently modelled as a set of curves analyzed by methods of Functional Data Analysis, or FDA -- see, e.g., \citet{Ramsay:2005}. For example, \citet{Morris:2007} view a child's activity level as a set of daily curves where metabolic activity is a function of time. Similarly, \citet{Fu:2017} view the dive profile of a southern elephant seal (\textit{Mirounga leonina}) as a set of dive curves whereby the amplitude and phase variation of the dive is used to classify dive types.
%

FDA was originally developed to process curves assumed to be independent replicates (i.e., there is no between-curve dependence), and within-curve fine-scale structure is not usually incorporated in FDA models. However, sets of curves often exhibit complex sequential dependencies both between curves and within curves, especially in the case of biologging data \citep{Barajas:2017}.
For example, dive profiles of marine animals show distinct dive ``types" that cluster together in time \citep{Tennessen:2019b} while simultaneously displaying bouts of short-term periodicity within dives \citep{Adam:2019}. Fine-scale periodicity within a larger process is also common in fields ranging from machine health \citep{Xin:2018,Lucero:2019} to speech recognition \citep{Juang:1991}.


Some FDA models account for between-curve dependence occurring when multiple curves arise from separate groups of individuals, but these models are inadequate when modelling certain types of dependence in time. For example, previous studies have used multilevel models with random effects to model variation between and within individuals in daily activity levels of children \citep{Morris:2007} or in menstrual cycles of adults \citep{Bromback:1998}. More recent work involving multilevel models includes that of \citet{Di:2009}, \citet{Crainiceanu:2009}, and \citet{Chen:2012}. However, these multilevel models are not appropriate in many biologging applications since they do not account for different curve types. Further, the first two papers do not account for temporal between-curve dependence and the third paper's model of temporal dependence is not appropriate in most biologging applications. 
In addition to multi-level models, FDA researchers have also used functional time series to model dependence in a sequence of curves. Functional time series extend the ideas of classic time series to model the evolution of one curve into the next \citep{Kokoszka:2018}, but they do not account for sequences of time-series curves whose distributions are determined by well-defined hidden states.

Traditional FDA techniques similarly fall short when modelling complicated within-curve 
data. In particular, within-curve structure is usually modelled by a generic smooth mean function and a covariance function \citep{Yao:2005} 
or with random regression \citep{Rice:2001}. 
However, time-series data exhibiting both sharp behavioural changes and periodic fine-scale structure are difficult to model with these classic FDA techniques.

To accommodate both temporal dependence and changes in curve type, we turn to the field of
animal movement modelling \citep{Hooten:2017}, where one of the most prevalent techniques of late is the hidden Markov model, or HMM \citep{Patterson:2017,McClintock:2020}. HMMs interpret animal movement data as arising from a sequence of unobserved behavioural states, allowing biologists to infer the underlying behaviour of an animal from sequential observations of its position. While ubiquitous in ecology literature, HMMs have seen little use in non-parametric functional modelling with a few notable exceptions. In particular, \citet{Langrock:2018} take a non-parametric approach to model the distributions of HMM observations with B-splines, 
but this approach does not account for certain types of temporal correlation.

While useful, HMMs alone are also not sufficient to model intricate fine-scale time-series data for three primary reasons.
Firstly, HMMs assume that subsequent observations are independent given an underlying hidden state process, but this is often not the case when observations are taken at extremely high frequencies. 
Several solutions have been proposed in the ecology literature such as the hidden movement Markov model \citep{Whoriskey:2016} and the conditionally auto-regressive hidden Markov model, or CarHMM \citep{Lawler:2019}. 
%
Secondly, classic HMMs fail to model simultaneous behavioural processes that occur at different time scales (i.e., both between and within curves). 
To address this issue, statistical ecologists have employed hierarchical hidden Markov models (HHMMs) \citep{Barajas:2017,Adam:2019}, which model both scales with conditionally dependent HMMs.
Thirdly, traditional HMMs, CarHMMs, and HHMMs cannot easily capture complicated dependence structures at short time scales. For example, \citet{Adam:2019} fail to capture fine-scale periodic swimming pattern of horn sharks (\textit{Heterodontus francisci}) using a traditional HHMM. \citet{Heerah:2017} successfully use Fourier analysis within an HMM to account for 
daily behavioural cycles in marine mammals, and Fourier analysis has previously been used with accelerometer data to explain animal behaviour \citep{Fehlmann:2017,Shorter:2017}. Thus, incorporating Fourier analysis within the structure of an HMM appears to be a promising approach to account for fine-scale periodic structures.

In this paper, we combine existing methods from statistical ecology literature in novel ways to account for complex, temporally dependent functional data. The resulting suite of methods makes up a ``tool box" that can be used to build arbitrarily complex hierarchical models to explain multi-scale functional and time-series data with intricate dependence structure.
We begin in Section 2 by describing HMMs as well as two variants, CarHMMs and HHMMs, and discuss how Fourier analysis can handle fine-scale dependence structures. We also show how these methods can be combined to analyze increasingly complex data. In Section 3 we fit several candidate models to data from a killer whale (\textit{Orcinus orca}) from the threatened northern resident population off the coast of British Columbia, Canada. Section 4 details a simulation study based on these candidate models, and in Section 5 we discuss our results.
\section{MODELS AND PARAMETER ESTIMATION}
\label{sec:models}
\pdfoutput=1


Consider a sequence of $T$ curves, where any particular curve $t$ is characterized by a curve-level (or coarse-scale) observation $Y_t$ as well as a sequence of $T^*_t$ within-curve (or fine-scale) observations $Y^*_{t}$. Namely, $Y^*_{t} \equiv \big\{Y^*_{t,1},\ldots,Y^*_{t,T^*_t}\big\}$ is made up fine-scale quantities derived from curve $t$ indexed using $t^*$. Both $Y_t$ and $Y^*_{t,t^*}$ can be either vectors or scalars. We call the sequence of coarse-scale observations $Y \equiv \big\{Y_1, \ldots, Y_T\big\}$ and the collection of all fine-scale observations $Y^* \equiv \big\{Y^*_1,\ldots,Y^*_T \big\}$. To develop our model for this data, we detail the structure of a traditional HMM followed by three variations which generalize its base structure. We then show how each of these generalized HMMs can be synthesized to form a wide variety of more complicated models.


\subsection{HMMs as a base structure}
\label{subsec:HMM}

Hidden Markov models describe state-switching Markovian processes in discrete time and are the core structure we use to model both $Y$ and $Y^*$. For simplicity we focus on $Y$ to introduce the model. An HMM is comprised of a sequence of unobserved states $X \equiv \big\{X_1, \ldots, X_T\big\}$ together with an observation sequence $Y \equiv \big\{Y_1, \ldots, Y_T\big\}$, where $X_t$ is associated with the observation $Y_t$. The $Y_t$'s are often referred to as ``emissions'' and the index $t$ typically refers to time. 
The $X_t$'s form a Markov chain and can take integer values between $1$ and $N$. Their distribution is governed by the distribution of the initial state $X_1$ and the $N \times N$ transition probability matrix $\Gamma$, where $\Gamma_{ij} = \Pr(X_{t+1} = j | X_t = i)$. 
We assume that $X_1$ follows the chain's stationary distribution, which is denoted by an $N$-dimensional row vector $\delta$, where
$\delta_i = \Pr(X_1 = i).$
A Markov chain's stationary distribution is determined by its probability transition matrix via $\delta = \delta \Gamma$ and $\sum_{i=1}^N \delta_i = 1$.
The distribution of an emission $Y_t$ conditioned on the corresponding hidden state $X_t$ does not depend upon any other observation or hidden state.
If $X_t=i$ then we denote the conditional density or probability mass function of $Y_t$ as $f^{(i)}(\cdot ; \theta^{(i)})$ or simply $f^{(i)}(\cdot)$, where $\theta^{(i)}$ is a state-dependent parameter describing the emission distribution.

Using observation emissions, denoted here as $y \equiv \{y_1,\ldots,y_T\}$, we can find the maximum likelihood estimates of the parameters $\Gamma$ and $\theta \equiv \{\theta^{(1)},\ldots,\theta^{(N)}\}$. The likelihood $\calL_{\text{HMM}}$ can be evaluated using the well-known \textit{forward algorithm} \citep{Zucchini:2016}:
$$\calL_{\text{HMM}}(\theta,\Gamma;y) = \delta P(y_1;\theta) \prod_{t=2}^T \Gamma P(y_t;\theta) \mathbf{1}_N,$$
where $\mathbf{1}_N$ is an $N$-dimensional column vector of ones and
$P(y_t;\theta)$ is an $N \times N$ diagonal matrix with $(i,i)^{th}$ entry $f^{(i)}(y_t; \theta^{(i)})$.

\begin{figure}[ht]
    \begin{subfigure}{\textwidth}
      \centering
      \includegraphics[width=4in]{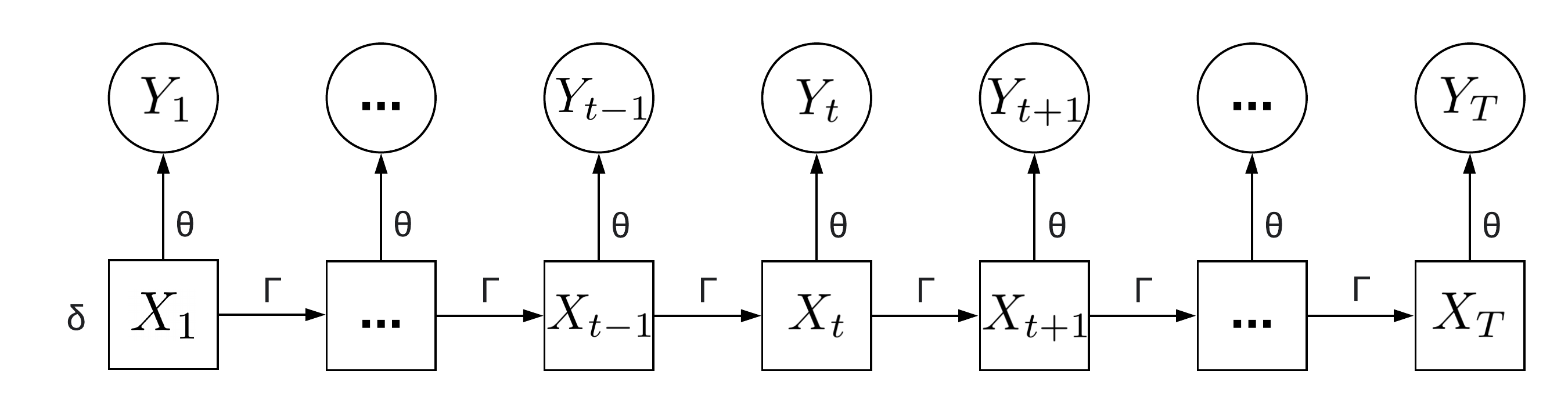}  
      \caption{Hidden Markov Model (\textbf{HMM})}
      \label{fig:HMM}
    \end{subfigure}
    \newline
    \begin{subfigure}{\textwidth}
      \centering
      \includegraphics[width=4in]{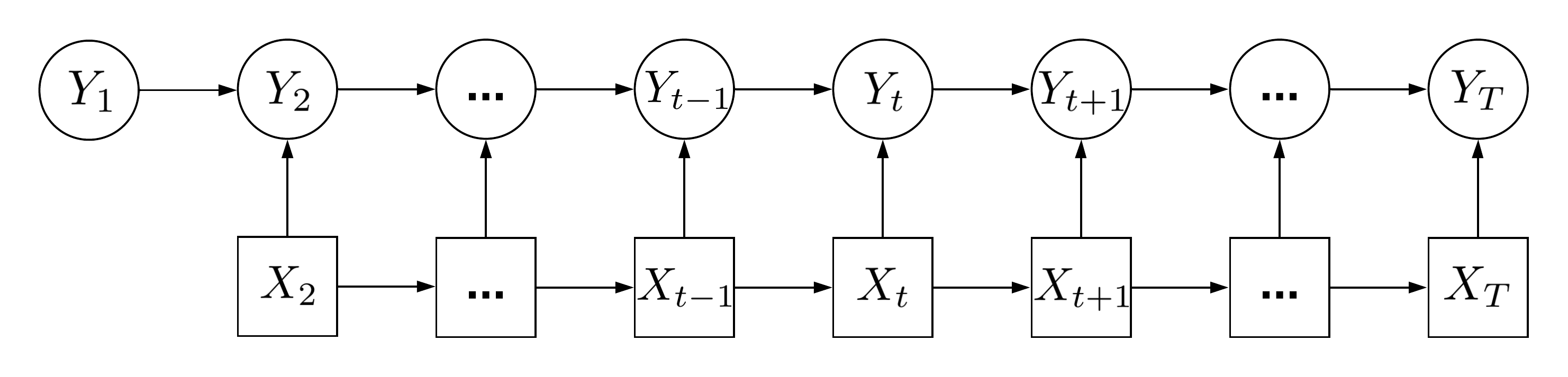}  
      \caption{Conditionally Autoregressive HMM (\textbf{CarHMM})}
      \label{fig:CarHMM}
    \end{subfigure}
    \newline
    \begin{subfigure}{\textwidth}
      \centering
      \includegraphics[width=4in]{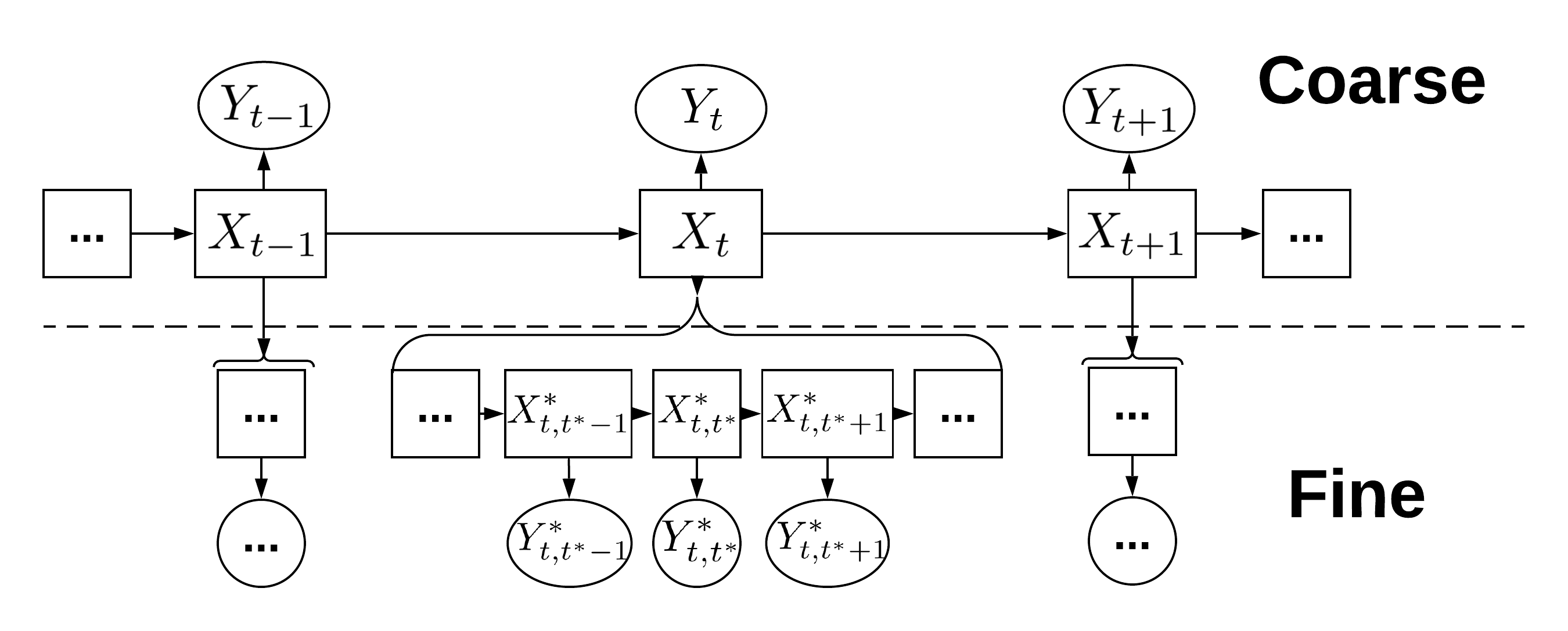}  
      \caption{Hierarchical HMM (\textbf{HHMM})}
      \label{fig:HHMM}
    \end{subfigure}
    \newline
    \begin{subfigure}{\textwidth}
      \centering
      \includegraphics[width=4in]{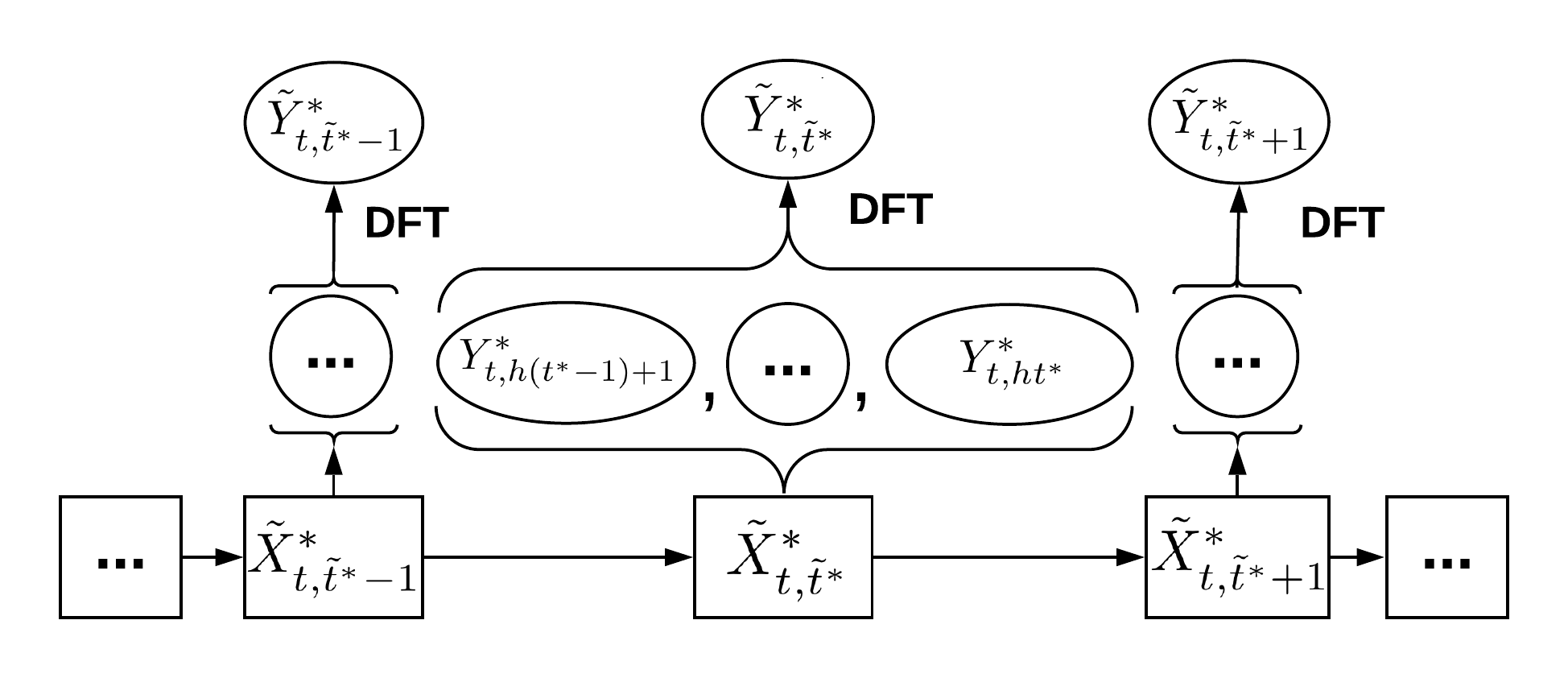}  
      \caption{HMM with Discrete Fourier Transform (\textbf{HMM-DFT})}
      \label{fig:HMM-DFT}
    \end{subfigure}
    \caption{Dependence structures of a standard HMM (a) and three HMM variants (b, c, and d). Hidden state sequences are denoted as $X$ on the coarse-scale and $X^*$ on the fine scale. Likewise, observations (or emissions) are denoted by $Y$ on the coarse scale and $Y^*$ on the fine scale. The HHMM shown here has a traditional HMM as both the coarse- and fine-scale model. In (d), observations are transformed using a moving window and denoted as $\Z$ with corresponding hidden states $\tilde X^*$.}
    \label{fig:models}
\end{figure}

Following \citet{Barajas:2017}, we reparameterize the $N \times N$ transition probability matrix $\Gamma$ such that the entries of the matrix are forced to be non-negative and the rows sum to 1:
\[
\Gamma_{ij} = \frac{\exp(\eta_{ij})}{\sum_{k=1}^N \exp(\eta_{ik})}, 
\]
where $i,j = 1,\ldots,N$ and $\eta_{ii}$ is set to zero for identifiability. This simplifies likelihood maximization by removing constraints in the optimization problem. For simplicity, we will continue to use $\Gamma$ in our notation, suppressing the reparameterization in terms of $\eta$. Figure \ref{fig:models}a represents the dependence structure of an HMM.

\subsection{Relaxing conditional independence with the CarHMM}
\label{subsec:CarHMM}

A conditionally auto-regressive hidden Markov model, or CarHMM \citep{Lawler:2019}, is a generalization of an HMM which explicitly models auto-correlation in the observation sequence beyond the correlation induced by the hidden state process. Like a traditional HMM, a CarHMM is made up of a Markov chain of unobserved states $X_1,\ldots,X_T$ that can take integer values between $1$ and $N$. A CarHMM also has a transition probability matrix $\Gamma$ and initial distribution $\delta$ equal to the stationary distribution of $\Gamma$. Unlike a traditional HMM, the CarHMM assumes that the distribution of $Y_t$ conditioned on $X_1,\ldots, X_T$ and $Y_1,\ldots, Y_{t-1}$ depends on both $X_t$ and $Y_{t-1}$ rather than only $X_t$. The first emission $Y_1$ is treated as a fixed initial value which does not depend upon $X_1$. We denote the conditional density or probability mass function of $Y_t$ given $Y_{t-1} = y_{t-1}$ and $X_t=i$ as $f^{(i)}(\cdot | y_{t-1}; \theta^{(i)})$ or simply $f^{(i)}(\cdot | y_{t-1})$. This model is highly general, as $f^{(i)}(\cdot | y_{t-1})$ can be any valid density or probability mass function that depends upon the parameters $\theta^{(i)}$ and the previous observation $Y_{t-1}$. 
As a concrete example, if $Y_t$ is a scalar, then one may assume that $Y_t$ given $X_t = i$ is Normally distributed with parameters $\theta^{(i)} = \{\mu^{(i)},\sigma^{(i)},\phi^{(i)}\}$, where:
\begin{align}
\label{eqn:carhmm}
\begin{split}
\mathbb{E}(Y_{t}|Y_{t-1} = y_{t-1},X_t=i) &= \phi^{(i)} ~ y_{t-1} ~+ ~(1-\phi^{(i)})  ~\mu^{(i)}, \\
\mathbb{V}(Y_t| Y_{t-1} = y_{t-1},X_t=i) &= (\sigma^{(i)})^2.
\end{split}
\end{align}
A CarHMM which follows Equation (\ref{eqn:carhmm}) can be viewed as a discrete time version of a state-switching Ornstein-Uhlenbeck process \citep{Michelot:2019}. This follows in the same way that an AR(1) process is the discrete-time version of a traditional Ornstein-Uhlenbeck process. 

As previously, the likelihood corresponding to a general CarHMM can be easily calculated using the forward algorithm. If $y$ is the sequence of observed emissions, then
\begin{equation*}
    \calL_{\text{CarHMM}}(\theta,\Gamma;y) = \delta \prod_{t=2}^T \Gamma P(y_t|y_{t-1};\theta) \mathbf{1}_N,
    \label{eqn:CarHMM_likelihood}
\end{equation*}
where
$P(y_t|y_{t-1};\theta)$ is an $N \times N$ diagonal matrix with $(i,i)^{th}$ entry equal to $f^{(i)}(y_t|y_{t-1}; \theta^{(i)})$. Figure \ref{fig:models}b shows a graphical representation of the dependence structure of a CarHMM. 

\subsection{Incorporating multiple scales with the HHMM}
\label{subsec:HHMM}

A hierarchical hidden Markov model, or HHMM, accounts for processes occurring simultaneously at different scales by modelling both the coarse-scale process and fine-scale process with either HMMs \citep{Barajas:2017,Adam:2019} or CarHMMs. The coarse-scale model is either an HMM and CarHMM as defined in Sections \ref{subsec:HMM} and \ref{subsec:CarHMM}, where $X_1, \ldots, X_T$ make up an unobserved Markov chain with $N$ possible states and $Y_1,\ldots, Y_T$ are the corresponding observations with state-dependent parameters $\theta^{(i)}$ for $i = 1,\ldots,N$.   
In the hierarchical setting, however, each state $X_t$ also emits another sequence of fine-scale unobserved states, $X_t^* \equiv \{X_{t,1}^*,\ldots, X_{t,T_t^*}\}$, which in turn emits a sequence of fine-scale observations $Y_t^* \equiv \{Y_{t,1}^*,\ldots, Y_{t,T_t^*}\}$. For each curve $t$, the fine-scale process $\{X_t^*, Y_t^*\}$ then follows another HMM (or CarHMM) whose parameters depend on the value of $X_t$. If $X_t = i$, then the components of $X_t^*$ make up a Markov chain with $N^{*(i)}$ possible states, an $N^{*(i)} \times N^{*(i)}$ transition probability matrix $\Gamma^{*(i)}$, and an initial distribution $\delta^{*(i)}$ which we assume is equal to the stationary distribution of the chain. The distribution of $Y^*_{t,t^*}$ given $Y^*_{t,t^*-1} = y^*_{t,t^*-1}$, $X^*_{t,t^*}=i^*$, and $X_t=i$ is governed by the parameter $\theta^{*(i,i^*)}$ and has density or probability mass function denoted $f^{*(i,i^*)}\left(\cdot|y^*_{t,t^*-1}; \theta^{*(i,i^*)}\right)$ or simply $f^{*(i,i^*)}(\cdot|y^*_{t,t^*-1})$. We denote the set of fine-scale emission parameters corresponding to $X_t=i$ as $\theta^{*(i)}=\big\{\theta^{*(i,1)}, \ldots, \theta^{*\left(i,N^{*(i)}\right)}\big\}$. In summary:
%
%
\begin{gather*}
    \{Y, X\} \text{ follows a (Car)HMM with } \Gamma \in \bbR^{N \times N}, \\
    (Y_t | Y_{t-1} = y_{t-1}, X_t = i) \text{ has density } f^{(i)}(\cdot|y_{t-1};\theta^{(i)}), \\
    \{Y^*_t,X^*_t | X_t = i\} \text{ follows a (Car)HMM with } \Gamma^{*(i)} \in \bbR^{N^{*(i)} \times N^{*(i)}}, \\
    (Y^*_{t,t^*} | Y^*_{t,t^*-1} = y^*_{t,t^*-1}, X^*_{t,t^*} = i^*, X_t = i) \text{ has density } f^{*(i,i^*)}(\cdot|y^{*}_{t,t^*-1};\theta^{(i,i^*)}).
\end{gather*}
Given the coarse-scale hidden state sequence $X$, the $T+1$ sets $\{X_1^*, Y_1^*\}, \ldots, \{X_T^*, Y_T^*\}$, and $\{Y_1,\ldots,Y_T\}$ are assumed to be independent of one another.
%
%

Forcing certain parameters to be shared can reduce complexity and increase interpretability of the HHMM. For example, in our killer whale case study (see Section \ref{sec:data}), we take $N^{*(i)} = N^*$ for all $i$. We also share the fine-scale emission parameters across the $N$ coarse-scale hidden states $\left( \text{i.e., } \theta^{*(1,i^*)} = \cdots = \theta^{*(N,i^*)} = \theta^{*(\cdot,i^*)} \text{ for all } i^* = 1, \ldots, N^* \right)$. Coarse-scale hidden states therefore differ only in their coarse-scale emission parameters $\theta^{(i)}$ and fine-scale probability transition matrices $\Gamma^{*(i)}$.

Due to the nested structure of the HHMM, the likelihood is easily calculated using the forward algorithm.
Let $y$ be the sequence of observed coarse-scale emissions and
$y^* \equiv \{y^*_1, \ldots,y^*_T\}$ be the collection of $T$ observed fine-scale emission vectors.
In addition, let $\theta^* \equiv \{\theta^{*(1)}, \ldots, \theta^{*(N)}\}$ denote the collection of all fine-scale emission parameters and $\Gamma^* \equiv \{\Gamma^{*(1)}, \ldots, \Gamma^{*(N)}\}$ denote the collection of all fine-scale transition probability matrices. The likelihood of the observed data is then
\begin{equation}
    \calL_{\text{HHMM}}(\theta,\theta^*,\Gamma,\Gamma^*;y,y^*) = \delta P_m(y_1,y_1^*;\theta,\theta^*,\Gamma^*) \prod_{t=2}^T \Gamma P_m(y_t,y_t^*|y_{t-1};\theta,\theta^*,\Gamma^*) \mathbf{1}_N
    \label{eqn:HHMM_likelihood}
\end{equation}
where $P_m(y_t,y_t^*|y_{t-1};\theta,\theta^*,\Gamma^*)$ is an $N \times N$ diagonal matrix whose exact structure depends upon the coarse- and fine-scale models. If the coarse-scale model is an HMM, $P_m(y_1,y_1^*;\theta,\theta^*,\Gamma^*)$ and $P_m(y_t,y_t^*|y_{t-1};\theta,\theta^*,\Gamma^*)$ for $t \geq 2$ both have $(i,i)^{th}$ entries equal to 
$f^{(i)}(y_t)\calL_{\text{fine}}\left(\theta^{*(i)},
\Gamma^{*(i)};y_t^*\right)$. 
If the coarse-scale model is a CarHMM, $P_m(y_1,y_1^*;\theta,\theta^*,\Gamma^*)$ has $(i,i)^{th}$ entry equal to $\calL_{\text{fine}}\left(\theta^{*(i)},
\Gamma^{*(i)};y_1^*\right)$ and $P_m(y_t,y_t^*|y_{t-1};\theta,\theta^*,\Gamma^*)$ for $t \geq 2$ has $(i,i)^{th}$ entry equal to $f^{(i)}(y_t|y_{t-1})\calL_{\text{fine}}\left(\theta^{*(i)},
\Gamma^{*(i)};y_t^*\right)$.
The fine-scale likelihood $\calL_{\text{fine}}$ corresponds to the likelihood of the fine-scale model, which can be either a CarHMM or an HMM. Figure \ref{fig:models}c graphically displays the dependence structure for an HHMM.
%

\subsection{Transforming fine-scale observations with the HMM-DFT}
\label{subsec:STFT}

In many applications where data are collected at high frequencies, intricate dependency structures arise within the fine-scale process that cannot be adequately modelled with the HMM-based models described thus far. 
To handle these additional fine-scale structures, we recommend replacing $Y_t^* = \{Y^*_{t,1},\ldots,Y^*_{t,T^*_t}\}$ with relevant statistics that summarize any non-Markovian behaviour. To maintain the temporal structure of the fine-scale process, local summary statistics can be calculated from a moving window over the elements of $Y_t^*$. Subject matter experts are often required to determine the specific summary statistics employed as well as the optimal window size and stride length of the moving window. Stride length refers to the distance between the first element of consecutive windows, so a stride length of $h$ indicates that the first window starts at $Y^*_{t,1}$, the second at $Y^*_{t,h+1}$, and so on. Larger stride lengths result in a loss of information but also reduce the dimension of the fine-scale process, which allows for faster model fitting. In addition, setting the stride length equal to the window size avoids artificial residual correlation arising from overlapping windows. For our case study, we use the discrete Fourier transform (DFT) of a moving forward window of width $h$ and stride $h$ across $Y^*_t$. Namely:
\begin{align}
    DFT\{Y^*_{t,t^*},\ldots, Y^*_{t,t^*+h-1}\}(k) = \sum_{n=0}^{h-1} Y^*_{t,t^*+n}\exp\left(-\frac{i 2\pi}{h} kn \right)
    \label{eq:DFTdef}
\end{align}
for $t^* = 1,h+1,2h+1,\ldots$ and $k = 0, 1, \ldots, h-1$ with $i = \sqrt{-1}$. If $Y^*_{t,t^*}$ is a vector then the DFT is taken component-wise. We omit the final window if $t^*+h-1$ exceeds $T^*_t$, denote the total number of windows as $\tilde T^*_t$, and index each window with $\tilde{t}^* = 1,\ldots,\tilde T^*_t$. Next, we calculate transformed observations $\Z_{t,\tilde{t}^*} \equiv \{\Zone_{t,\tilde{t}^*},\Ztwo_{t,\tilde{t}^*}\}$:
\begin{equation}
    \label{eqn:z}
    \Zone_{t,\tilde{t}^*} \equiv \frac{1}{h}\sum_{n=1}^{h}Y^*_{t,h(\tilde{t}^*-1)+n}, \quad \Ztwo_{t,\tilde{t}^*} \equiv \sum_{k=1}^{\tilde{\omega}}\Big|\Big|DFT\{Y^*_{t,h(\tilde{t}^*-1)+1},\ldots, Y^*_{t,h\tilde{t}^*}\}(k)\Big|\Big|^2,
\end{equation}
where $\tilde{\omega} \leq h-1$ is a problem-specific tuning parameter corresponding to the maximum recorded frequency within each window. In words, $\Zone_{t,\tilde{t}^*}$ is the average value of $Y^*_t$ within window $\tilde{t}^*$ and $\Ztwo_{t,\tilde{t}^*}$ is the squared two-norm of the component of the window that can be attributed to frequencies between one and $\tilde{\omega}$ periods per window. 
More intuitively, $\Ztwo_{t,\tilde{t}^*}$ corresponds to the ``wiggliness" of the fine-scale data within curve $t$ and window $\tilde{t}^*$.

When performing this transformation, the fine-scale HMM (or CarHMM) must be redefined since $\Z_t$ exists on a coarser scale than $Y^*_t$ itself. As a result, there are only $\tilde{T}_t^*$ hidden states associated with $\Z_t = \big\{\Z_{t,1},\ldots, \Z_{t,\tilde{T}^*_t}\big\}$, which we denote as $\tilde{X}^*_t = \big\{\tilde{X}^*_{t,1},\ldots, \tilde{X}^*_{t,\tilde{T}^*_t}\big\}$. The fine-scale probability transition matrices  $\Gamma^{*(i)}$ and probability density functions $f^{*(i,i^*)}$ are then applied directly to $\tilde{X}^*_t$ and $\Z_t$ instead of $X^*_t$ and $Y^*_t$. The likelihood of this model is therefore identical to that of the original HMM or CarHMM as defined in Sections \ref{subsec:HMM} and \ref{subsec:CarHMM}, but $Y^*$ is replaced with $\Z$ and $X^*$ is replaced with $\tilde{X^*}$. To clearly differentiate models, we refer to an HMM with $\Z$ as observations and $\tilde{X^*}$ as hidden states as an HMM-DFT. Figure \ref{fig:models}d displays the dependence structure of an HMM-DFT.

\subsection{Generalized hierarchical Markov models}

Traditional HHMMs treat both the coarse-scale and the fine-scale processes as realizations of an HMM or CarHMM. 
However, the fine-scale observations for a particular dive $Y^*_t$ can be modelled using a large variety of parametric models which admit easy-to-compute likelihoods or penalized likelihoods. As such, the fine-scale HMM likelihood term $\calL_{\text{fine}}$ in Equation (\ref{eqn:HHMM_likelihood}) can be replaced by the likelihood of a general fine-scale model whose parameters depend upon the coarse-scale hidden state.
%
Possible candidates for the fine-scale model include any of the models described in the previous subsections in addition to many others not described here. 
For example, \citet{Bebbington:2007} and \citet{Borchers:2013} investigate data sets with count onsets as observations, so they use variations of a Poisson process as their fine-scale model. If the fine-scale model is a simple Poisson process, then this approach is equivalent to a Markov-modulated Poisson process \citep{Fischer:1993}.
The fine-scale process can also be modelled as in \citep{Langrock:2018}, who use B-splines to model the emission distribution of an HMM. This non-parametric approach uses a penalized likelihood term which can easily replace the usual fine-scale likelihood term in Equation (\ref{eqn:HHMM_likelihood}). 
Another class of fine-scale models is the set of continuous time methods such as a continuous-time HMM (CTHMM) \citep{Liu:2015} or a state-switching Ornstein-Uhlenbeck process \citep{Michelot:2019}. 
A continuous-time HMM may be appropriate if observations are not equi-spaced in time \citep{Liu:2015}. \citet{Xu:2018} model high-frequency biologging accelerometer data of individuals by incorporating a CTHMM into a hierarchical model similar to ours. However, they assume that individuals are partitioned into subgroups a priori whereas we use an HMM to infer the coarse-scale hidden states.

These examples include a few of many fine-scale models that can act as initial building blocks in a practitioner's toolbox to construct increasingly complex hierarchical models. 
A myriad of possible models can be built using this framework, but these models can quickly become complicated and computationally expensive to fit. Therefore, models should be constructed with care to achieve an adequate fit of the data while avoiding over-fitting and high computational costs.

%
\section{KILLER WHALE CASE STUDY}
\label{sec:data}
\pdfoutput=1


To illustrate the process of constructing a model using these building blocks, we analyze the dive behaviour of a northern resident killer whale in Queen Charlotte Sound off the coast of British Columbia, Canada, and construct several candidate models to categorize and describe its diving behaviour.

Understanding animal behaviour is important for conservation efforts, as environmental changes caused by anthropogenic activity can directly impact animal behaviour \citep{Sutherland:1998}. HMMs have been used to understand how diving behaviours of various species are affected by disturbances (e.g., \citet{DeRuiter:2017} and \citet{Isojunno:2017}). For killer whales, we are interested in categorizing different diving behaviours to identify potential foraging dives. Northern resident killer whales feed almost exclusively on calorie-rich Chinook salmon (\textit{Oncorhynchus tshawytscha}) \citep{Ford:2006}, which typically occur deeper and are less numerous than smaller types of salmon \citep{Ford:2009}. Northern resident killer whales therefore must expend significant amounts of energy to capture Chinook \citep{Williams:2009,Noren:2011,Wright:2017}. 
Acceleration data can be used to estimate an animal's energy expenditure \citep{Green:2009,Wilson:2019}, but the animal's behavioural state must be accounted for in order to obtain accurate estimates \citep{Dot:2016}. Therefore, understanding both the behavioural state of the killer whale and the distribution of acceleration within each behavioural state is needed to determine the true energetic requirements of the animal.

\subsection{Data collection and preprocessing}

\begin{figure}[tb!]
	\centering
	\includegraphics[width=5.25in]{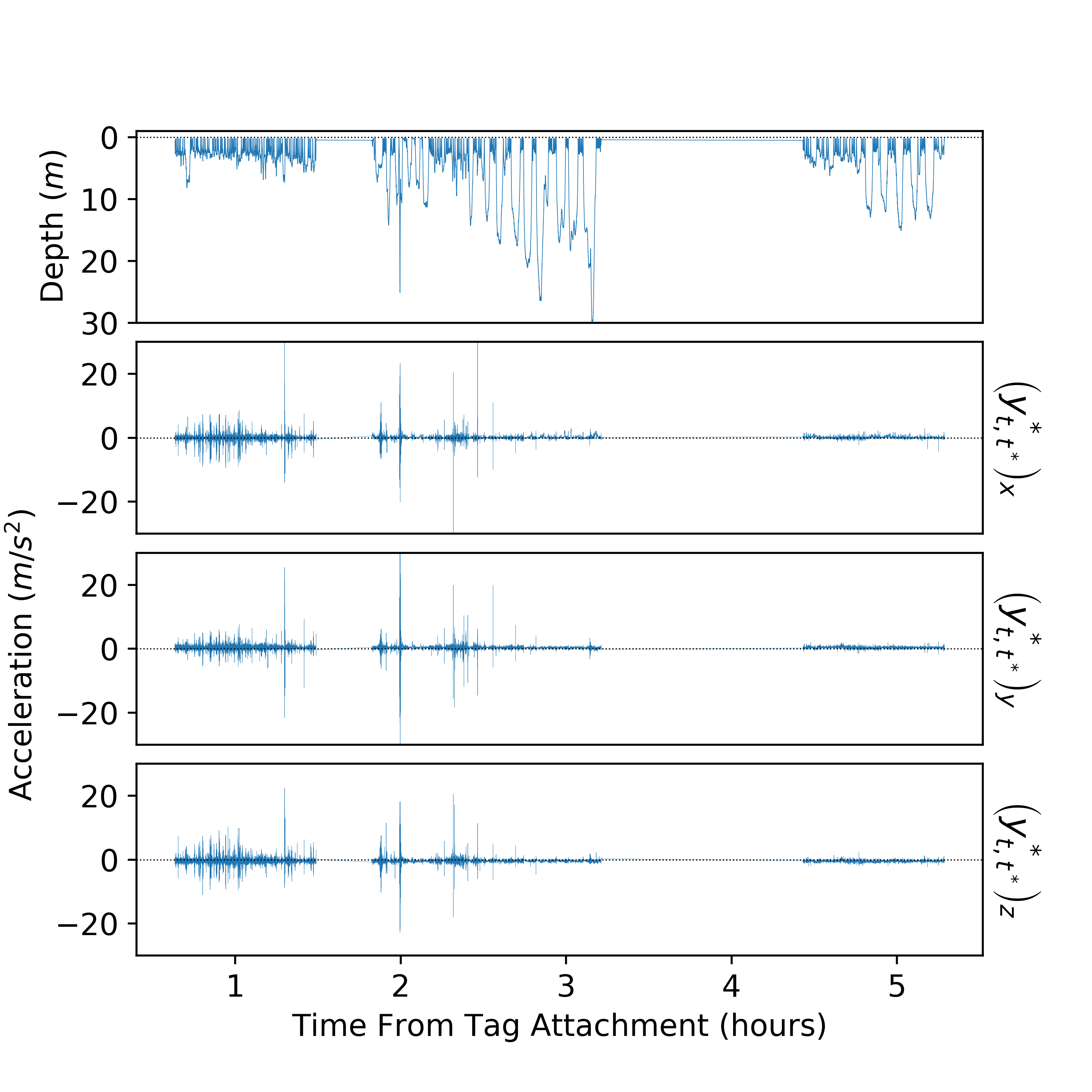}
	\caption{Dive depth (top panel) and three-dimensional acceleration (bottom three panels) from a killer whale over approximately 5 hours. An exact physical interpretation of each of component acceleration is difficult due to variations in tag orientation. There are data gaps occurring from around $1.5$ to $1.8$ hours and from around $3.2$ to $4.5$ hours. Both data gaps were excluded from the analysis.}
	\label{fig:data}
\end{figure}

The data we use were collected on September 2, 2019 from 12:49 pm to 6:06 pm PDT and consist of depth and acceleration over time. Observations were collected at a rate of 50 Hz using a CATs biologger (Customizable Animal Tracking Solutions, {\em{www.cats.is}}). Acceleration was measured in three dimensions, which together represent the complete range of movement of an animal (forward/backward, upward/downward, and right/left). Tri-axial acceleration readings are common in these types of tags and are often used to infer animal behaviour such as foraging \citep{Cade:2017,Fehlmann:2017,Wright:2017}. The act of attaching and detaching the tag caused anomalous behaviour before 1:20 pm and after 6:00 pm, so observations taken during these time periods are ignored. There were also periods of time when the tag failed to record observations, resulting in data gaps between 2:25 pm and 2:37 pm and between 4:07 pm and 5:07 pm. To preprocess the data, we smooth the depth and acceleration curves by taking a moving average within a window of $1/10^{th}$ of a second. We then define a killer whale ``dive" as any continuous interval of data that occurs below 0.5 meters in depth and lasts for at least 10 seconds. Data are preprocessed in part with the \textit{divebomb} package in Python \citep{Nunes:2018}. The preprocessed data contain a total of 267 dives, all of which are displayed in Figure \ref{fig:data}. Each dive is treated as one curve, and the sequence of dives makes up the coarse-scale process. Specifically, the sequence of observed coarse-scale observations $y = \big\{y_1,\ldots,y_{267}\big\}$ is a time series of dive durations in seconds. For dive $t$, the fine-scale observations are contained in $y^*_{t} \equiv \big\{y^*_{t,1},\ldots,y^*_{t,T^*_t} \big\}$, which is a sequence containing the within-dive acceleration data in units of meters per second squared. The collection of all acceleration data is denoted as $y^* = \big\{y^*_1,\ldots,y^*_{267}\big\}$.

\subsection{Model definition and selection}
\label{subsec:model_selection}

Defining a suitable model to describe this killer whale kinematic data involves selecting an appropriate number of hidden states, model structure, and emission distributions for both the coarse- and fine-scale observations.

We do not use information criteria to select the number of dive types $N$ since these metrics tend to overestimate the number of behavioural states in biological processes \citep{Pohle:2017}. We instead plot the duration of each dive versus the duration of the dive preceding it ($y_t$ versus $y_{t-1}$ for $t \geq 2$). This type of visualization is known as a lag plot. If the emission distributions of the hidden states are well-separated, a lag plot should reveal $N$ distinct patterns, where each pattern corresponds to one dive type \citep{Lawler:2019}. This is unfortunately not the case for our killer whale data, as there is one cluster of data centred at approximately $y_t = y_{t-1} = 30$ seconds. However, longer dives appear to be characterized by bouts of less ``wiggly" behaviour in the acceleration data compared to shorter dives, so we choose $N = 2$ to differentiate these dive types. The absence of a more principled method to select $N$ highlights the importance of model validation techniques in lieu of information criteria (see Section \ref{subsec:model_validation}).
Prior to fitting the model, lag plots reveal no significant auto-correlation between dive duration observations (see Figure 1 of the supplementary material), and visual inspection shows no obvious complicated dependence. Therefore, we select a simple HMM to model the coarse-scale process since neither a CarHMM nor a moving-window transformation is called for.
Given that dive $t$ is of type $i$, we assume that the dive duration $Y_t$ follows a Gamma distribution with unknown parameters $\mu^{(i)}$ and $\sigma^{(i)}$:
$$\mathbb{E}(Y_t|X_t = i) = \mu^{(i)}, \qquad \mathbb{V}(Y_t|X_t = i) = \left(\sigma^{(i)}\right)^2.$$
This is consistent with previous studies, including that of \citet{Barajas:2017}. 

We then select a model corresponding to the fine-scale observations of acceleration. Similarly to the coarse model, we rely on lag plots and visual inspection to select $N^*=3$ subdive states. Although $N^*$ is selected heuristically, we test the validity of this model in Section \ref{subsec:model_validation}.
In contrast to the coarse-scale observations, the fine-scale acceleration data exhibit significant sinusoidal behaviour. Thus, we transform each fine-scale observation sequence $y_t^*$ into $\z_t$ using Equation (\ref{eqn:z}) with a window size of $h=100$ (two seconds) and a maximum frequency of $\tilde{\omega}=10$ (5 Hz). 
We then have that $\z_{t,\tilde t^*} = \{\zone_{t,\tilde t^*},\ztwo_{t,\tilde t^*}\}$, where $\zone_{t,\tilde t^*}$ is a three dimensional vector of component-wise average acceleration and $\ztwo_{t,\tilde t^*}$ is a scalar describing the ``wiggliness" of a particular window. Even after transforming the raw acceleration data, there is still strong auto-correlation within each component of $\zone_{t,\tilde t^*}$ prior to fitting the model (see Figure 1 of the supplementary material). Therefore, we choose a CarHMM as defined in Section 2.2 as the fine-scale model.

We then select the specific emission distribution of $\Z_{t,\tilde t^*}$. First, we assume that $\Ztwo_{t,\tilde t^*}$ and all three components of $\Zone_{t,\tilde t^*}$ are independent of one another when conditioned on the dive types and subdive states. To reduce model complexity, we also assume that the three sets of fine-scale emission parameters are shared across the two dive types $\left(\theta^{*(1,i^*)} = \theta^{*(2,i^*)} \equiv \theta^{*(\cdot,i^*)} \text{ for } i^* = 1,2,3\right)$. This implies that the subdive states within dive type 1 have the same interpretation as those within dive type 2.
To specify the emission distribution of $\Zone_{t,\tilde t^*}$, consider the sequence $\big\{\Zone_{t,1},\ldots,\Zone_{t,T^*_t}\big\}$ for a particular dive $t$. We assume that each of the three components of this sequence are Normally distributed as in Equation (\ref{eqn:carhmm}) and all components are independent of one another when conditioned on the subdive states $\big\{\tilde X_{t,1},\ldots,\tilde X_{t,\tilde T^*_t}\big\}$. Each component is assumed to have its own mean and variance parameters, but all components share the same auto-correlation parameter. Thus the distribution of $\Zone_{t,\tilde t^*}$ given $\Zone_{t,\tilde t^*-1}$ and $X^*_{t,\tilde t^*} = i^*$ has parameters $\mathbf{\mu}_A^{*(\cdot,i^*)} \in \mathbb{R}^3$, $\mathbf{\sigma}_A^{*(\cdot,i^*)} \in \mathbb{R}^3$, and $\phi_A^{*(\cdot,i^*)} \in [0,1]$.
To specify the emission distribution of $\Ztwo_{t,\tilde t^*}$, we assume that given $\tilde X^*_{t,\tilde t^*} = i^*$, $\Ztwo_{t,\tilde t^*}$ follows a Gamma distribution parameterized by its mean $\mu_W^{*(\cdot,i^*)}$ and standard deviation $\sigma_W^{*(\cdot,i^*)}$. In addition, $\Ztwo_{t,1},\ldots,\Ztwo_{t,T^*_t}$ are assumed to be independent of one another given the subdive state sequence $\big\{\tilde X_{t,1}, \ldots, \tilde X_{t,T_t^*}\big\}$. We do not include $\Ztwo_{t,\tilde t^*-1}$ in the distribution of $\Ztwo_{t,\tilde t^*}$ because the auto-correlation evident from the lag plot is not severe and may be explained by subsequent observations occurring within the same subdive state. 
%

In total, the parameters to estimate are
\begin{gather*}
    \Gamma, \qquad \Gamma^{*} = \{\Gamma^{*(1)},\Gamma^{*(2)}\} \qquad \text{(probability transition matrices)}, \\
    \theta = \{\mu^{(1)},\sigma^{(1)},\mu^{(2)},\sigma^{(2)}\} \qquad \text{($Y$ emission parameters), and} \\
    \theta^* = \{\theta^{*(\cdot,1)},\theta^{*(\cdot,2)},\theta^{*(\cdot,3)}\}  \qquad \text{($\Z$ emission parameters), where} \\
    \theta^{*(\cdot,i^*)} =  \{\mu_A^{*(\cdot,i^*)},\sigma_A^{*(\cdot,i^*)},\phi_A^{*(\cdot,i^*)},\mu_W^{*(\cdot,i^*)},\sigma_W^{*(\cdot,i^*)}\}.
\end{gather*}
Recall that $\theta^{*(\cdot,i^*)}$ is the set of parameters describing the distribution of $\Z_{t,\tilde t^*}$ conditioned on $\tilde X^*_{t,\tilde t^*} = i^*$. 
We refer to this final model as the \textbf{CarHHMM-DFT} since it includes a CarHMM, hierarchical HMM, and DFT-based transformation. The likelihood of this model is easily calculated using the forward algorithm and can be maximized with respect to the parameters above (see the appendix for details). Figure \ref{fig:CarHHMM-DFT} shows the dependence structure of the full CarHHMM-DFT.

\begin{figure}[tb!]
	\centering
	\includegraphics[width=5in]{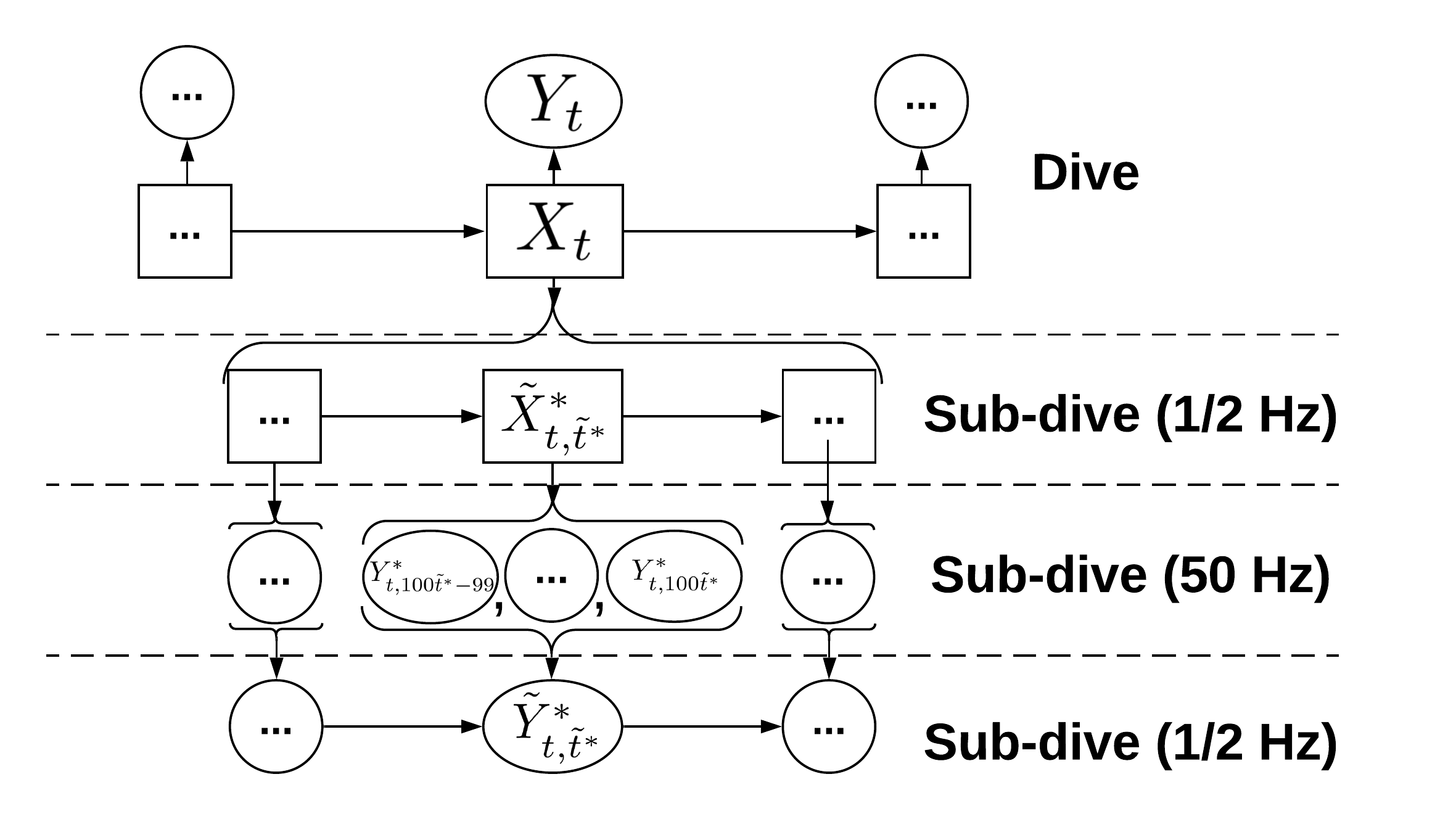}
	\caption{Graphical representation of the conditionally auto-regressive hierarchical hidden Markov model with discrete Fourier transform (\textbf{CarHHMM-DFT}) used in the simulation and case study. The type of dive $t$ is denoted by $X_t$ and $Y_t$ represents the associated dive duration. The raw acceleration vector associated with dive $t$ and time stamp $t^*$ is denoted by $Y^*_{t,t^*}$. The subdive state of the killer whale during dive $t$ and window $\tilde t^*$ is denoted as $\tilde X^*_{t,\tilde t^*}$, and the corresponding transformed observation is denoted by $\Z_{t,\tilde t^*}$.}
	\label{fig:CarHHMM-DFT}
\end{figure}

In addition to the CarHHMM-DFT, we consider three variations for comparison. As in the full model, each of the following models assume that all components of $\Z_{t,\tilde t^*}$ are conditionally independent of one another given the dive types and subdive states:
\begin{enumerate}
    \item An \textbf{HHMM-DFT}, which models the coarse-scale observations with an HMM and transforms the fine-scale observations using Equation (\ref{eqn:z}), but models $\Z_{t,\tilde t^*}$ as emissions from a simple HMM rather than a CarHMM.
    \item A \textbf{CarHHMM}, which models the coarse-scale observations with an HMM, transforms the fine-scale observations using Equation (\ref{eqn:z}), and models $\Zone_{t,\tilde t^*}$ as emissions of a CarHMM. However, the ``wiggliness"  $\Ztwo_{t,\tilde t^*}$ is omitted from this model altogether.
    \item A \textbf{CarHMM-DFT}, which models the coarse-scale observations as an independent and identically distributed sequence of dives, transforms the fine-scale observations using Equation (\ref{eqn:z}), and models $\Z_{t,\tilde t^*}$ as emissions of a CarHMM. This model assumes that there is only one dive type.
\end{enumerate}
Each of the three candidate models above leaves out one important aspect of the full CarHHMM-DFT: the HHMM-DFT assumes there is no auto-correlation between fine-scale observations, the CarHHMM does not incorporate ``wiggliness" $\big(\Ztwo_{t,\tilde t^*}\big)$, and the CarHMM-DFT lacks a hierarchical structure and thus does not distinguish between dive types.

\subsection{Case study results}

To illustrate an application of this method and compare the candidate models, we fit all four models to the data shown in Figure \ref{fig:data}. We first report the results from the full CarHHMM-DFT in detail and assess the quality of the fit. We then compare these results with those from the other candidate models.

\begin{figure}[tb!]
    \begin{subfigure}{\textwidth}
    	\centering
    	\includegraphics[width=4in]{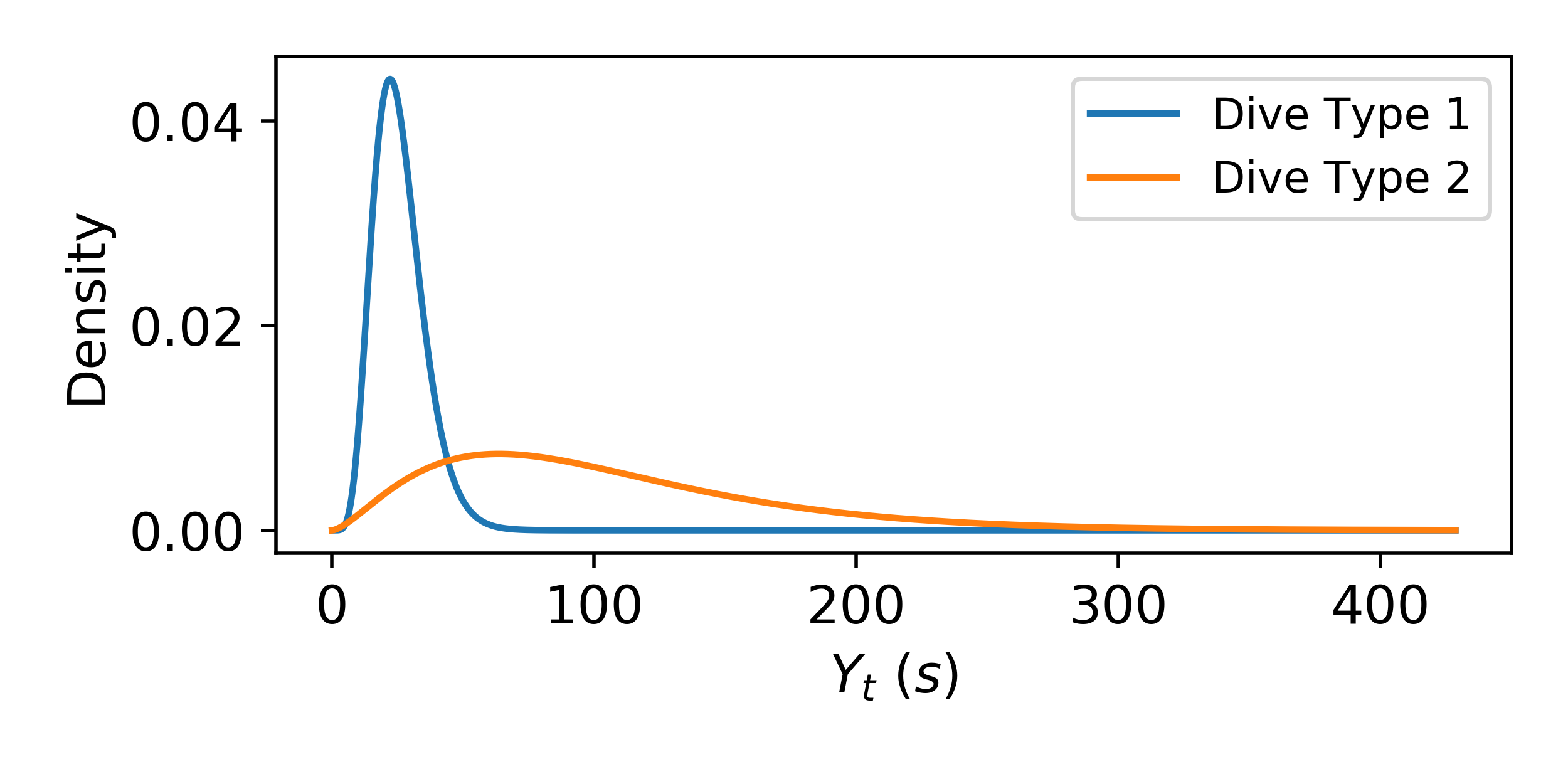}
    	\caption{Estimated Gamma probability density functions of a killer whale's dive duration ($Y_t$) corresponding to dive types 1 and 2.}
    	\label{fig:coarse_emis}
    \end{subfigure}
    \newline
    \begin{subfigure}{\textwidth}
    	\centering
    	\includegraphics[width=4in]{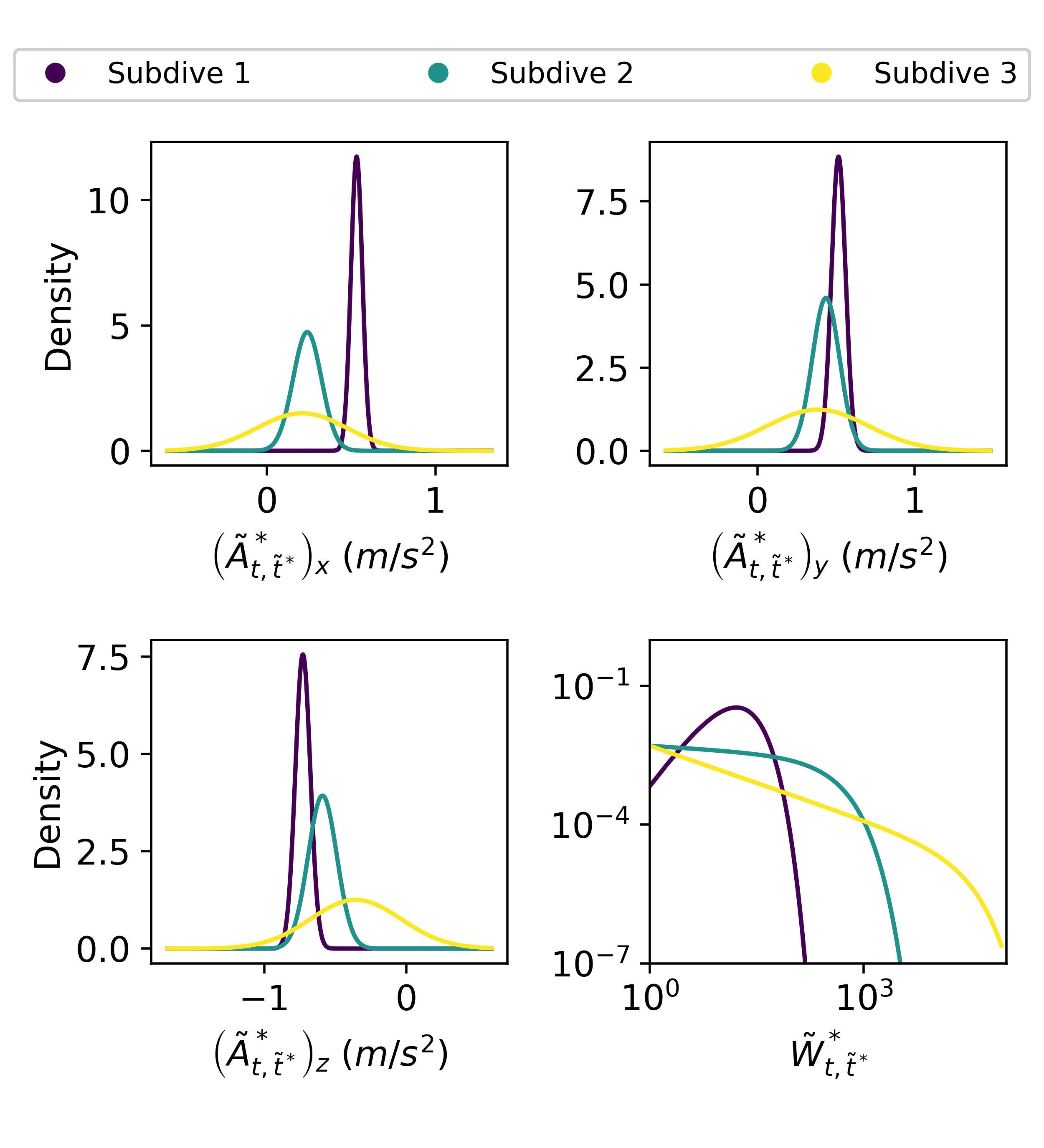}
    	\caption{Estimated Normal conditional densities of the three components of a killer whale's acceleration $\left(\Zone_{t,\tilde t^*}|\Zone_{t,\tilde t^*-1} = \mu_A^{*(\cdot,i^*)}\right)$ plotted on a linear scale and estimated Gamma probability density of the killer whale's wiggliness $\left(\Ztwo_{t,\tilde t^*}\right)$ plotted on a log-log scale. Density functions correspond to subdive states $i^* = 1,2,$ and $3$. Density functions corresponding to acceleration are conditioned on $\Zone_{t,\tilde t^*-1} = \mu_A^{*(\cdot,i^*)}$ because the density function of $\Zone_{t,\tilde t^*}$ itself depends on $\Zone_{t,\tilde t^*-1}$.}
    	\label{fig:fine_emis}
    \end{subfigure}
    \caption{Estimated probability density functions for coarse-scale and fine-scale emissions corresponding to observations of killer whale behaviour. Densities are estimated by fitting the CarHHMM-DFT to the case study data (see Table \ref{table:emis_dists_CarHHMM-DFT}).}
    \label{fig:emis}
\end{figure}

\begin{table}[tb!]
\centering
\caption{Estimates with standard errors for the parameters of the distributions of dive duration $\big(Y_t\big)$, acceleration $\big(\Zone_{t,\tilde t^*}\big)$, and wiggliness $\big(\Ztwo_{t,\tilde t^*}\big)$ of the killer whale kinematic data using the full CarHHMM-DFT. The parentheses refer to standard errors estimated using the observed information matrix.}
\begin{tabular}{ccccc}
    \multirow{2}{*}{Feature}                                                       & \multirow{2}{*}{Dive / Subdive Type} & \multicolumn{3}{c}{Parameter Estimate (Standard Error)}                      \\
                                                                                   &                                      & $\hat \mu$        & $\hat \sigma$   & $\hat \phi$       \\ \hline
    \multirow{2}{*}{Dive Duration $(s)$ -- $Y_t$}                                  & 1                                    & $25.68 \ (0.60)$   & $9.57\ (0.51)$   & ---               \\
                                                                                   & 2                                    & $104.6 \ (9.4)$    & $64.7\ (7.5)$    & ---               \\ \hline
    \multirow{3}{*}{$x$-Acc. $(m/s^2)$ -- $\left(\Zone_{t,\tilde t^*}\right)_x$}   & 1                                    & $0.020 \ (0.042)$  & $0.034\ (0.001)$ & $0.976\ (0.007)$ \\
                                                                                   & 2                                    & $0.244 \ (0.013)$  & $0.079\ (0.001)$ & $0.886\ (0.005)$ \\
                                                                                   & 3                                    & $0.218 \ (0.028)$  & $0.265\ (0.007)$ & $0.626\ (0.029)$ \\ \hline
    \multirow{3}{*}{$y$-Acc. $(m/s^2)$ -- $\left(\Zone_{t,\tilde t^*}\right)_y$}   & 1                                    & $0.469 \ (0.052)$  & $0.044\ (0.001)$ & $0.976\ (0.009)$ \\
                                                                                   & 2                                    & $0.436 \ (0.014)$  & $0.082\ (0.001)$ & $0.886\ (0.012)$ \\
                                                                                   & 3                                    & $0.384 \ (0.033)$  & $0.321\ (0.009)$ & $0.626\ (0.034)$ \\ \hline
    \multirow{3}{*}{$z$-Acc. $(m/s^2)$ -- $\left(\Zone_{t,\tilde t^*}\right)_z$}   & 1                                    & $-0.683\ (0.061)$  & $0.052\ (0.001)$ & $0.976\ (0.005)$ \\
                                                                                   & 2                                    & $-0.593\ (0.016)$  & $0.096\ (0.001)$ & $0.886\ (0.009)$ \\
                                                                                   & 3                                    & $-0.366\ (0.033)$  & $0.317\ (0.009)$ & $0.626\ (0.033)$ \\ \hline
    \multirow{3}{*}{Wiggliness - $\Ztwo_{t,\tilde t^*}$}                           & 1                                    & $23.34 \ (0.29)$   & $12.95\ (0.27)$  & ---               \\
                                                                                   & 2                                    & $301.2 \ (3.2)$    & $330.1\ (4.2)$   & ---               \\
                                                                                   & 3                                    & $10200 \ (210)$    & $15300\ (350)$   & ---               \\ \hline
    \end{tabular}
    \label{table:emis_dists_CarHHMM-DFT}
\end{table}

The coarse-scale parameter estimates suggest that the killer whale has at least two distinct dive behaviours (see Table \ref{table:emis_dists_CarHHMM-DFT} and Figure \ref{fig:coarse_emis}). 
Dive type 1 corresponds to shorter, shallower dives which likely reflect resting, travelling, and to a lesser extent searching for prey.
Dive type 2 is longer, deeper, and may be associated with behaviours such as hunting  \citep{Tennessen:2019b}, but it is unclear whether any of the dives in this study are successful foraging dives. No dive in this data set has a maximum depth deeper than 30 meters and a study by \citet{Wright:2017} of killer whales in Johnstone Strait found that most prey captures occur at depths deeper than 100 meters. However, the killer whale studied here was tagged north of Johnstone Strait in Queen Charlotte Sound, and significant numbers of fish have been observed to be caught near the surface (Fortune and Trites, unpublished data). Dive type 2 could also be associated with behaviours such as socializing that can take place several meters below the surface \citep{Tennessen:2019b}.

\begin{figure}[t]
	\centering
	\includegraphics[width=5in]{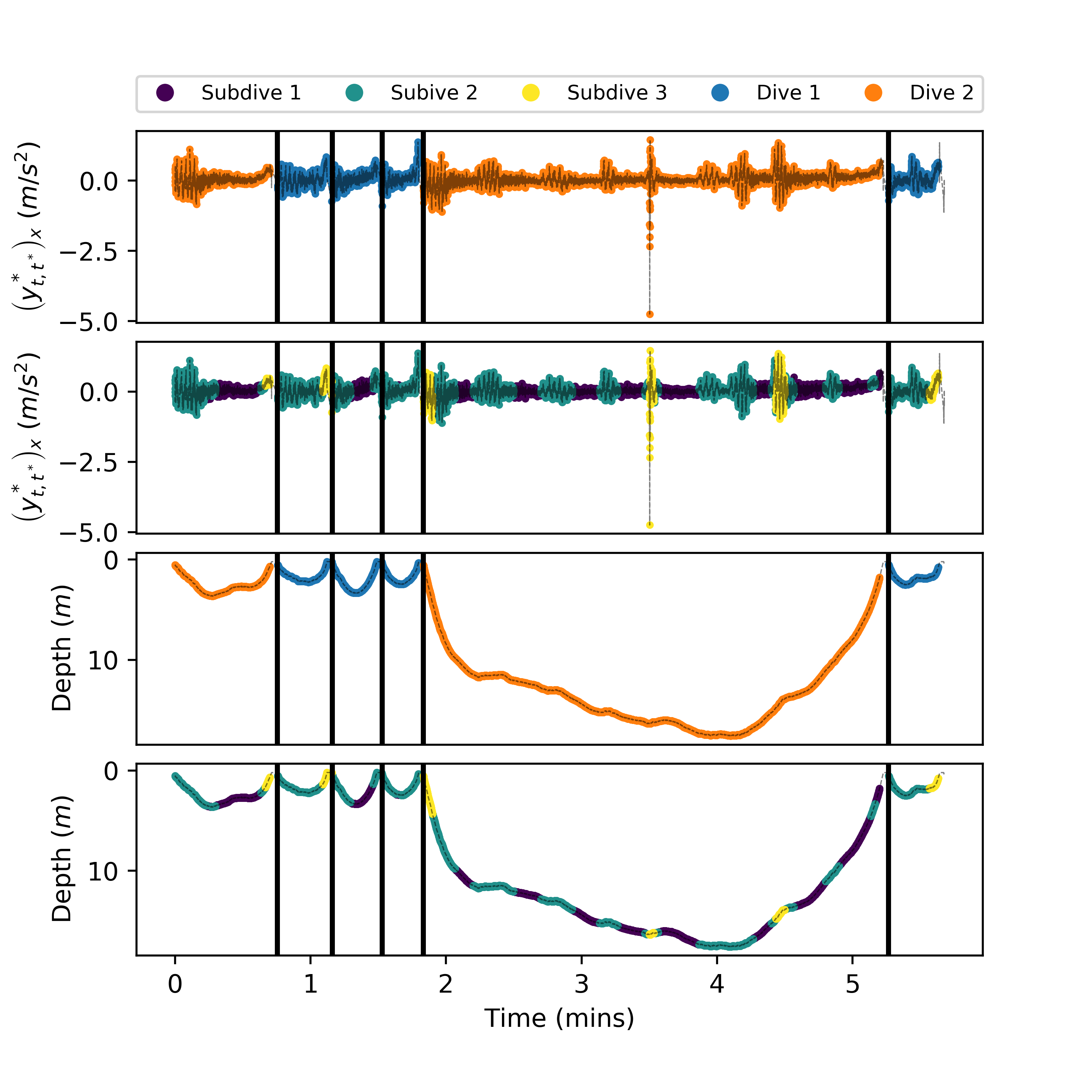}
	\caption{The $x$--component of acceleration $\left(y^*_{t,t^*}\right)_x$ (top two panels) and dive depth (bottom two panels) of a northern resident killer whale for a sequence of six selected dives. Each panel is partitioned into dives by vertical black lines. The curve colours in the first and third panels correspond to the estimated dive types while the curve colours of the second and fourth panels correspond the estimated subdive states. Both the dive types and subdive states are estimated by fitting the CarHHMM-DFT to the data and performing the forward-backward algorithm to determine the hidden state with the highest probability.}
	\label{fig:labeled_dives}
\end{figure}

The means of ``wiggliness" $\big(\Ztwo_{t,\tilde t^*}\big)$ associated with each subdive state are separated by an order of magnitude (see Table \ref{table:emis_dists_CarHHMM-DFT} and Figure \ref{fig:fine_emis}). 
Subdive state 1 has the smallest mean corresponding to $\Ztwo_{t,\tilde t^*}$ and the smallest variance corresponding to $\Zone_{t,\tilde t^*}$. It also has the highest auto-correlation in $\Zone_{t,\tilde t^*}$. This implies less overall activity and more consistent acceleration compared to the other subdive states. 
Subdive state 2 has a mean ``wiggliness" $\big(\Ztwo_{t,\tilde t^*}\big)$ one order of magnitude higher than subdive state 1 and its acceleration has about twice the variance compared to subdive state 1. The auto-correlation of acceleration is also slightly lower than subdive state 1. We therefore hypothesize that subdive state 2 corresponds to fluking (active swimming), as strong sinusoidal behaviour in acceleration is characteristic of this behaviour in marine mammals \citep{Simon:2012}.
Finally, the mean of $\Ztwo_{t,\tilde t^*}$ and variance of $\Zone_{t,\tilde t^*}$ in subdive state 3 are both much higher than in the other two states, and the auto-correlation of $\Zone_{t,\tilde t^*}$ is also much lower. This corresponds to vigorous swimming activity, especially as the killer whale begins or ends a dive (see Figure \ref{fig:labeled_dives}). 

The estimated probability transition matrices and associated stationary distributions on the coarse scale are
$$\hat \Gamma = \begin{pmatrix} 
0.788 & 0.212 \\
0.809 & 0.191
\end{pmatrix} \text{ and }$$
$$\hat \delta = \begin{pmatrix} 0.792 & 0.208 \end{pmatrix}$$
for the transitions between dives. The estimated probability transition matrices and stationary distributions on the fine scale are 
$$\hat \Gamma^{*(1)} = \begin{pmatrix} 
0.679 & 0.321 & 0.000 \\
0.038 & 0.904 & 0.058 \\
0.000 & 0.232 & 0.768
\end{pmatrix}, \qquad 
\hat \Gamma^{*(2)} = \begin{pmatrix} 
0.859 & 0.141 & 0.000 \\
0.114 & 0.841 & 0.045 \\
0.000 & 0.216 & 0.784
\end{pmatrix},$$
$$\hat \delta^{*(1)} = \begin{pmatrix} 0.087 & 0.731 & 0.182 \end{pmatrix}, \enspace \text{and} \enspace \hat \delta^{*(2)} = \begin{pmatrix} 0.401 & 0.496 & 0.103 \end{pmatrix}$$
for dive types 1 and 2.
In summary, about 79\% of dives are short dives of type 1, and
the whale performs an average of 4.72 short type 1 dives before switching to dive type 2 and an average of 1.24 longer type 2 dives before switching back to dive type 1. This finding is consistent with those of \citet{Tennessen:2019b} and \citet{Williams:2009}, both of whom describe common bouts of short resting dives before a killer whale performs a longer, more energy-intensive deep dive. 
Further, this killer whale is in the less active subdive state 1 40\% of the time during a dive of type 1 compared to only only 9\% of the time during a dive of type 2. Less active swimming behaviour is consistent with the need for marine mammals to conserve energy when diving to depth and holding their breath for long periods \citep{Williams:1999,Hastie:2006}. Figure \ref{fig:labeled_dives} shows the decoded dive behaviour of six selected dives, and Section 1.3 of the supplementary material also shows the probability of each dive type and subdive state given the data and the fitted model.

\subsection{Model validation}
\label{subsec:model_validation}

\begin{figure}[tb!]
\centering
    \begin{subfigure}{0.4\textwidth}
    	\centering
    	\includegraphics[width=2.25in]{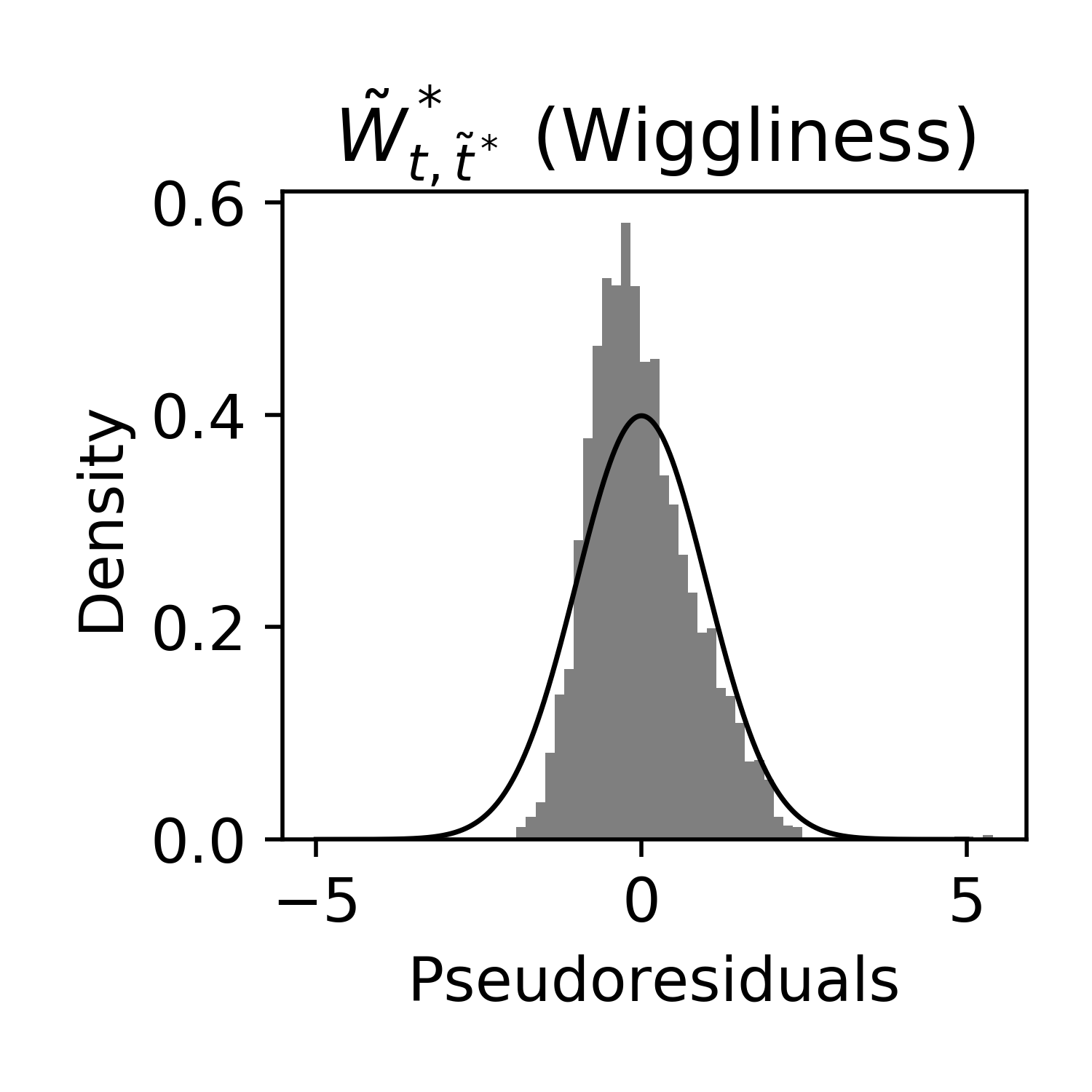}
    	\caption{Histogram of pseudoresiduals of $\tilde W^*_{t,\tilde t^*}$}
    	\label{fig:pseudoresids}
    \end{subfigure}
    \begin{subfigure}{0.4\textwidth}
    	\centering
    	\includegraphics[width=2.25in]{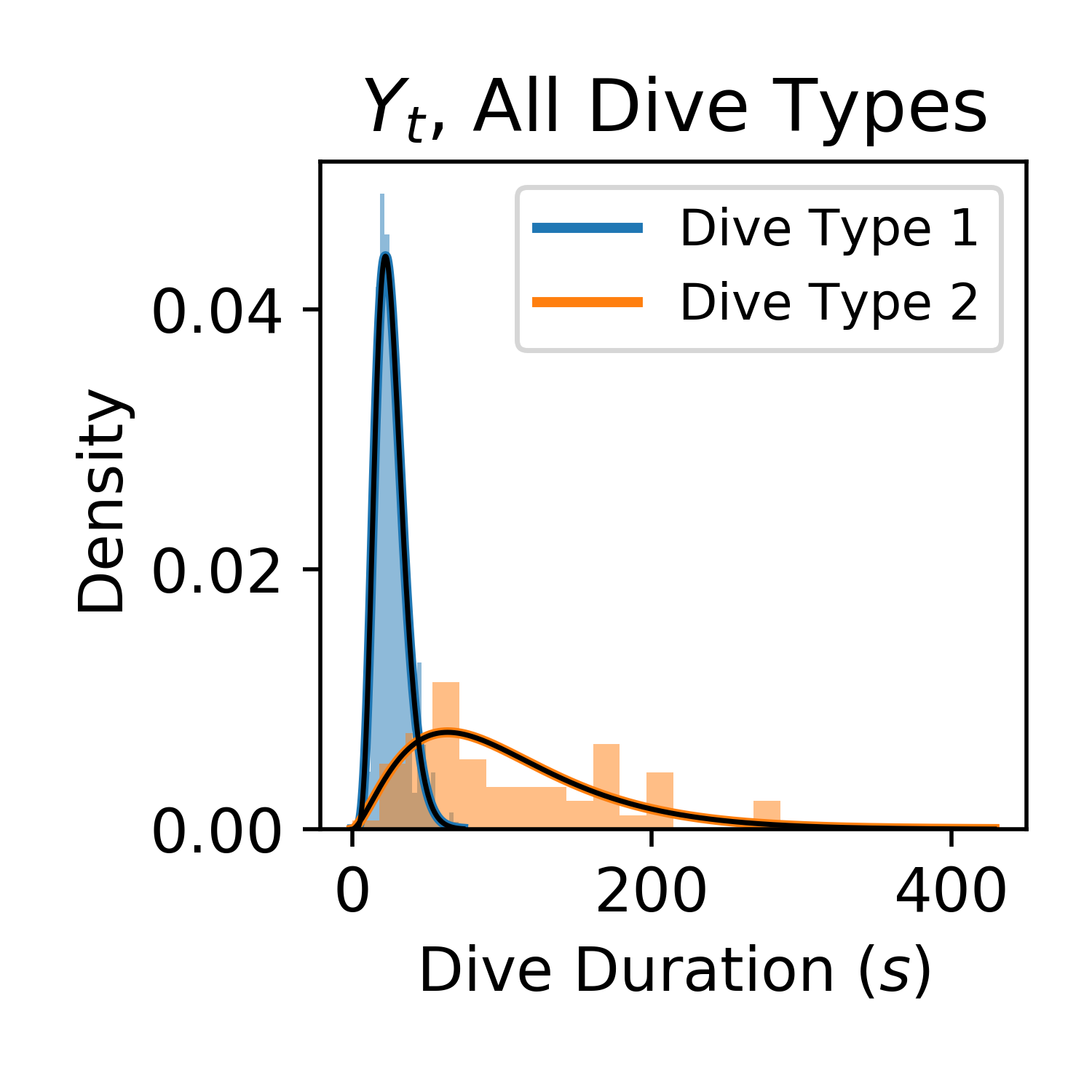}
    	\caption{Empirical distribution of $Y_t$}
    	\label{fig:empirical_dist}
    \end{subfigure}
    \caption{Pseudoresiduals of wiggliness ($\Phi^{-1} \big(\Pr(\Ztwo_{t,\tilde t^*} < \ztwo_{t,\tilde t^*}|Y,\tilde Y^* / \{\Z_{t,\tilde t^*}\}) \big)$, left) plotted over a standard normal density as well as a weighted empirical distribution of dive duration ($Y_t$, right) plotted over the corresponding fitted Gamma distributions. Both plots are generated by fitting the CarHHMM-DFT to the killer whale case study data and performing the forward-backward algorithm.}
    \label{fig:model_checking}
\end{figure}

We use two visual tools to evaluate the CarHHMM-DFT: pseudoresidual plots and empirical histograms. The pseudoresidual of a coarse-scale observation $y_t$ is equal to $\Phi^{-1} \left(\Pr(Y_t < y_t|\{Y_1,\ldots,Y_T,\Z_1,\ldots,\Z_T\}/\{Y_t\}) \right)$ and the pseudoresidual of a fine-scale observation $\z_{t,\tilde t^*}$ is $\Phi^{-1} \left(\Pr(\Z_{t,t^*} < \z_{t,\tilde t^*}|\{Y_1,\ldots,Y_T,\Z_1,\ldots,\Z_T\}/\{\Z_{t,\tilde t}\}) \right)$, where $\Phi$ is the cumulative distribution function of a standard Normal distribution. If the model is correct, then all pseudoresiduals are independent and follow a standard Normal distribution. Histograms of the pseudoresiduals mostly support that the CarHHMM-DFT is well-specified. One exception is $\Ztwo_{t,\tilde t^*}$, whose pseudoresiduals are noticeably right-skewed (see Figure \ref{fig:pseudoresids}). This implies that the true distribution of $\Ztwo_{t,\tilde t^*}$ may follow a heavier-tailed distribution than the Gamma distribution used in the case study. See Sections 1.4 through 1.6 of the supplementary material for pseudoresidual plots corresponding to all observations and models.

We also plot histograms of dive duration corresponding to each dive type in Figure \ref{fig:empirical_dist}. Each observation of dive duration is weighted by the estimated probability that it corresponds to a particular dive type as decoded by the \textit{forward-backward algorithm} \citep{Zucchini:2016}. This procedure results in two histograms -- one corresponding to dive type 1 and another corresponding to dive type 2. Each histogram is then plotted together with the corresponding emission distribution estimated by the CarHHMM-DFT. Analogous histograms corresponding to the fine-scale observations are contained in Section 1 of the supplementary material. Our results mostly show that the CarHHMM-DFT explains the data well, but there are some exceptions. In histograms corresponding to subdive state 3, $\Ztwo_{t,\tilde t^*}$ is right-skewed and $\Zone_{t,\tilde t^*}$ has heavier tails compared to a normal distribution, indicating the existence of rare events corresponding to exceptionally sudden changes in acceleration of the killer whale. These outliers are potential subjects for future study and may indicate biologically relevant phenomena such as prey capture \citep{Tennessen:2019a}.

\subsection{Comparison with candidate models}

The HHMM-DFT, which ignores auto-correlation in acceleration, decodes dive types and subdive states similarly to the CarHHMM-DFT, but it is less likely to categorize the behaviour at the beginning and end of dives as subdive state 3 (see Figures 4 and 5 of the supplementary material). In addition, for all three components of each of $\sigma_A^{*(\cdot,1)}$, $\sigma_A^{*(\cdot,2)}$, and $\sigma_A^{*(\cdot,3)}$, the HHMM-DFT produces estimates which are approximately $50$ to $100$ percent larger than those of the CarHHMM-DFT. The estimated uncertainties of the three components of each of $\hat \mu_A^{*(\cdot,1)}$, $\hat \mu_A^{*(\cdot,2)}$, and $\hat \mu_A^{*(\cdot,3)}$ are also less than half of those for the CarHHMM-DFT (see Tables 1 and 2 of the supplementary material). This suggests that including auto-correlation in the model significantly affects parameter estimates. Further, the pseudoresiduals of the HHMM-DFT are noticeably light-tailed and do not follow a standard normal distribution (see Figure 15 of the supplementary material). These findings suggest that the HHMM-DFT is a significantly worse fit to these data than the full CarHHMM-DFT.

The CarHHMM does not model the ``wiggliness" of the acceleration data, so it regularly fails to pick up obvious behavioural changes corresponding to the periodicity shown in Figure \ref{fig:labeled_dives} (see Figure 6 of the supplementary material). These results essentially disqualify the CarHHMM as a viable model for this data set. The pseudoresiduals of acceleration are also light-tailed relative to a Normal distribution (see Figure 16 of the supplementary material).

Finally, the CarHMM-DFT, which lacks a hierarchical structure, produces fine-scale parameter estimates and subdive state estimates similar to those of the CarHHMM-DFT. However, its lack of hierarchical structure means that it fails to differentiate between short and long dives. This model therefore does not infer the dive-level Markov chain or the relationship between the dive and subdive levels. In particular, the CarHMM-DFT does not indicate that the whale is more likely to be in subdive state 1 when engaged in longer dives compared to shorter dives. 

A more complete set of results for each of the candidate models is presented in Section 1 of the supplementary material.

\section{SIMULATION STUDY}
\label{sec:sim_study}
\pdfoutput=1

We perform a simulation study based on data generated from the full CarHHMM-DFT as defined in Section \ref{subsec:model_selection} to evaluate each candidate model when the ground-truth is known. The parameters used to generate the data are based on those estimated in the case study (see Table \ref{table:emis_dists_CarHHMM-DFT}), with slight modifications made for simplicity. In particular, we set the number of subdive states to $N^*=2$ and $\Zone_{t,\tilde t^*}$ to a scalar instead of a three dimensional vector. We then fit all four models to the simulated data. Metrics used to evaluate each model include decoding accuracy of hidden states, bias in parameter estimates, empirical standard errors of parameter estimates, and fitting times. To assess the accuracy of uncertainty estimates, we also compare the empirical standard errors of a given model's parameter estimates with the standard errors estimated using the inverse of the observed Fisher information.

\subsection{Simulation procedure}
\label{subsec:data_simulation}

We generate 500 independent training data sets using the CarHHMM-DFT as a generative model. Each training data set consists of a sequence of 100 curves which we call a sequence of killer whale dives. Each dive can be one of $N=2$ dive types based on a Markov chain with probability transition matrix
$$\Gamma = \begin{pmatrix} 0.79 & 0.21 \\ 0.81 & 0.19 \end{pmatrix}.$$
Dive duration is Gamma distributed and the coarse-scale emission parameters are 
\[\mu^{(1)} = 25.7s, \enspace \sigma^{(1)} = 9.6s, \enspace \mu^{(2)} = 104.6s, \enspace \sigma^{(2)} = 64.7s.\]
After generating the dive durations for all 100 dives in a data set, dive $t$ is broken into a sequence of $\tilde T^*_t = \lfloor Y_t/2 \rfloor$ two-second windows, where the last $Y_t - 2 \tilde T^*_t$ seconds of each simulated dive are ignored. Each two-second segment is assigned one of $N^*=2$ behaviours according to a fine-scale Markov chain $\tilde X^*_t \equiv \big\{\tilde X^*_{t,1}, \ldots, \tilde X^*_{t,\tilde T^*_t} \big\}$ with probability transition matrices
\[\Gamma^{*(1)} = \begin{pmatrix} 0.68 & 0.32 \\ 0.05 & 0.95 \end{pmatrix} \quad \text{ and } \quad \Gamma^{*(2)} = \begin{pmatrix} 0.86 & 0.14 \\ 0.15 & 0.85 \end{pmatrix}\]
for dive types 1 and 2, respectively.
Instead of generating the raw observations $Y^*_{t,t^*}$, we directly simulate the fine-scale transformed observations $\Z_{t,\tilde t^*} = \big\{\Zone_{t,\tilde t^*}, \Ztwo_{t,\tilde t^*}\big\}$. Recall from Section \ref{subsec:model_selection} that we must specify the mean, standard deviation, and auto-correlation parameters corresponding to $\big\{\Zone_{t,1},\ldots,\Zone_{t,\tilde T_t^*}\big\}$ as well as the mean and standard deviation parameters corresponding to $\big\{\Ztwo_{t,1},\ldots,\Ztwo_{t,\tilde T_t^*}\big\}$. We select the following parameters in line with the results from the case study:
%
\begin{gather*}
    \mu_A^{*(\cdot,1)} = 0.0 s, \enspace \sigma_A^{*(\cdot,1)} = 0.034s, \enspace \phi_A^{*(\cdot,1)} = 0.98, \\
    \mu_A^{*(\cdot,2)} = 0.0 s, \enspace \sigma_A^{*(\cdot,2)} = 0.079s, \enspace \phi_A^{*(\cdot,2)} = 0.87, \\
    \mu_W^{*(\cdot,1)} = 23.3, \quad \sigma_W^{*(\cdot,1)} = 13.0, \\
    \mu_W^{*(\cdot,2)} = 301.2, \quad \sigma_W^{*(\cdot,2)} = 330.1.
\end{gather*}
It is not possible to uniquely reconstruct the raw accelerometer data $Y^*$ from $\Z$ alone, but we describe one possible mapping from $\Z$ to $Y^*$ in the appendix. Figure 21 of the supplementary material shows one realization of $\Z$ for five dives of one simulated data set along with the corresponding reconstructed realization of $Y^*$. 

The two simulated dive types differ in that dives of type 1 are much shorter on average (26 seconds) than dives of type 2 (105 seconds). The two simulated subdive states differ primarily due to $\mu_W^*$ and $\sigma_W^*$ since both are much higher for subdive state 2 than for subdive state 1. These larger parameter values correspond to much more vigorous and variable periodic behaviour in the acceleration data. 

\begin{table}[t]
\centering
\caption{Average decoding accuracies and training times for all models used to categorize dive type and subdive state in the simulation study. Each of the four models was fit to 500 training data sets comprised of 100 simulated dives and tested on test data sets also comprised of 100 simulated dives. Reported values are averages, and $\pm$ refers to the sample standard deviation across the 500 data sets. Rows labelled Both/Both correspond to overall average decoding accuracy.}
\begin{tabular}{ccccccc}
Model                       & \multicolumn{1}{c}{Train Time (min)} & \multicolumn{1}{c}{Dive Type} & \multicolumn{1}{c}{Subdive State} & \multicolumn{1}{c}{Dive Accuracy} & \multicolumn{1}{c}{Subdive Accuracy}  \\ \hline
\multirow{5}{*}{CarHMM-DFT} & \multirow{5}{*}{$70 \pm 11$}   & Both                          & Both                             & -------------                     & $0.93 \pm 0.01$                       \\
                            &                                    & 1                             & 1                                & \multirow{2}{*}{-------------}    & $0.79 \pm 0.04$                       \\ 
                            &                                    & 1                             & 2                                &                                   & $0.96 \pm 0.01$                       \\ 
                            &                                    & 2                             & 1                                & \multirow{2}{*}{-------------}    & $0.92 \pm 0.01$                       \\ 
                            &                                    & 2                             & 2                                &                                   & $0.94 \pm 0.01$                       \\ \hline 
\multirow{5}{*}{HHMM-DFT}   & \multirow{5}{*}{$209 \pm 51$}   & Both                          & Both                             & $0.94 \pm 0.04$                   & $0.88 \pm 0.04$                       \\
                            &                                    & 1                             & 1                                & \multirow{2}{*}{$0.97 \pm 0.03$}    & $0.63 \pm 0.12$                       \\ 
                            &                                    & 1                             & 2                                &                                   & $0.96 \pm 0.01$                       \\ 
                            &                                    & 2                             & 1                                & \multirow{2}{*}{$0.85 \pm 0.10$}    & $0.79 \pm 0.16$                       \\ 
                            &                                    & 2                             & 2                                &                                   & $0.92 \pm 0.03$                       \\ \hline
\multirow{5}{*}{CarHHMM}    & \multirow{5}{*}{$236 \pm 52$}   & Both                          & Both                             & $0.87 \pm 0.17$                   & $0.74 \pm 0.05$                       \\
                            &                                    & 1                             & 1                                & \multirow{2}{*}{$0.87 \pm 0.21$}    & $0.77 \pm 0.21$                       \\ 
                            &                                    & 1                             & 2                                &                                   & $0.71 \pm 0.09$                       \\ 
                            &                                    & 2                             & 1                                & \multirow{2}{*}{$0.82 \pm 0.14$}    & $0.87 \pm 0.23$                       \\ 
                            &                                    & 2                             & 2                                &                                   & $0.66 \pm 0.10$                       \\ \hline
\multirow{5}{*}{CarHHMM-DFT}& \multirow{5}{*}{$132 \pm 40$}   & Both                          & Both                             & $0.94 \pm 0.04$                   & $0.93 \pm 0.01$                       \\
                            &                                    & 1                             & 1                                & \multirow{2}{*}{$0.96 \pm 0.03$}    & $0.76 \pm 0.04$                       \\ 
                            &                                    & 1                             & 2                                &                                   & $0.96 \pm 0.01$                       \\ 
                            &                                    & 2                             & 1                                & \multirow{2}{*}{$0.87 \pm 0.09$}    & $0.93 \pm 0.01$                       \\ 
                            &                                    & 2                             & 2                                &                                   & $0.93 \pm 0.01$                       \\ \hline
\end{tabular}
\label{table:accuracy}
\end{table}

We calculate maximum likelihood estimates $\{\hat \theta, \hat \Gamma, \hat \theta^*, \hat \Gamma^*\}$ for all four candidate models for each of the 500 data sets using the Cedar Compute Canada cluster with 1 CPU and 4 GB of dedicated memory per data set. For each of the 500 training data sets, we simulate a test data set to assess how well each model predicts the hidden states, as follows.
Each test data set consists of a sequence of 100 dives and is created from the generative model with true parameters $\{\theta, \Gamma, \theta^*, \Gamma^*\}$.
To assess the coarse-scale hidden state prediction, we estimate $p_t(i|y,\z) \equiv \Pr(X_t=i|Y=y,\Z=\z)$, $i=1,2$, $t=1,\ldots,100$ using the test-set observations $(y,\z)$ and training-set maximum likelihood estimates. These estimates are found using the {\em{forward-backward}} algorithm \citep{Zucchini:2016}. We compare these estimated conditional probabilities to $\{x_1,\ldots,x_{100}\}$, the true coarse-scale state realizations in the test data, by calculating the {\em{average dive decoding accuracy}} for a single training/test data set pair, $\sum_{t=1}^{100} \hat{p}_t(x_t|y,\z)/100$. We then report the average of these over the 500 training/test data set pairs. 
Analogously, to assess prediction of the fine-scale states, we estimate $p^*_{t,\tilde t^*}(i^*|y,\z) \equiv \Pr(\tilde X^*_{t,\tilde t^*}=i|Y=y,\Z=\z)$, $i^*=1,2$, $\tilde t^* = 1,\ldots,\tilde T^*_t$, $t=1,\ldots,100$, using the test-set observations, the training-set maximum likelihood estimates, and the forward-backward algorithm. Denoting the true fine-scale state realizations from the test data set and dive $t$ as $\{\tilde x^*_{t,1},\ldots,\tilde x^*_{t,\tilde T^*_t}\}$, we define the overall \textit{average subdive decoding accuracy} as the average value of $\hat{p}^*_{t,\tilde t^*}(\tilde x^*_{t,\tilde t^*}|y,\z)$ across all simulated test data sets, dives, and windows. The conditional probabilities are estimated according to one of the four models under study, using the maximum likelihood estimates from the training data set in conjunction with the forward-backward algorithm.

\subsection{Simulation results}

\begin{figure}[t]
    \centering
    \includegraphics[width=4.5in]{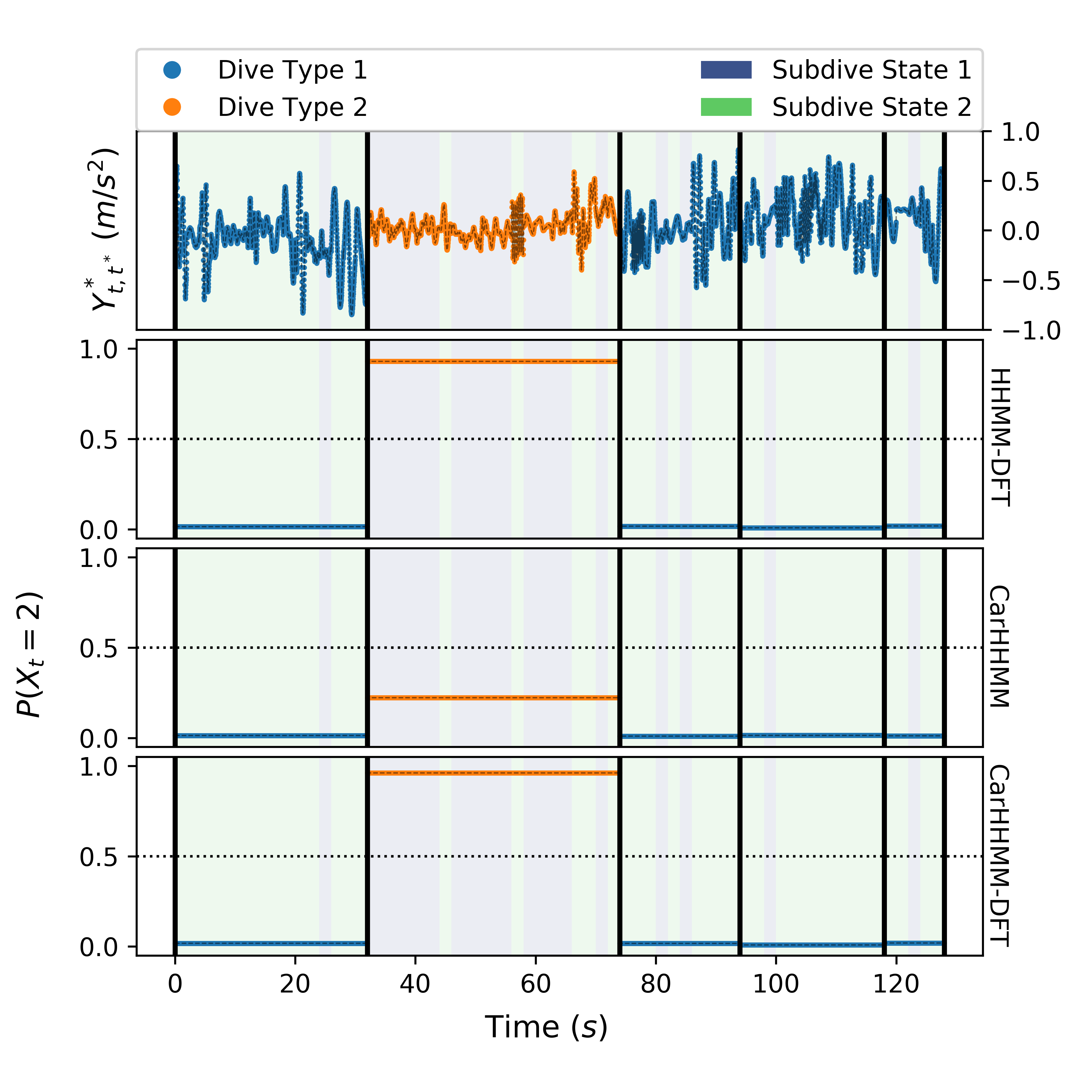}
    \caption{Estimated probabilities that each dive is of type 2 for five selected dives of a simulated data set of killer whale dive behaviour. Each panel is partitioned into dives by vertical black lines. The colour of the curve corresponds to the true dive type while the colour of the background corresponds to the true subdive state. The CarHMM-DFT is omitted because it assumes that there is only one dive type.}
    \label{fig:acc_coarse}
\end{figure}

\begin{figure}[t]
    \centering
    \includegraphics[width=4.5in]{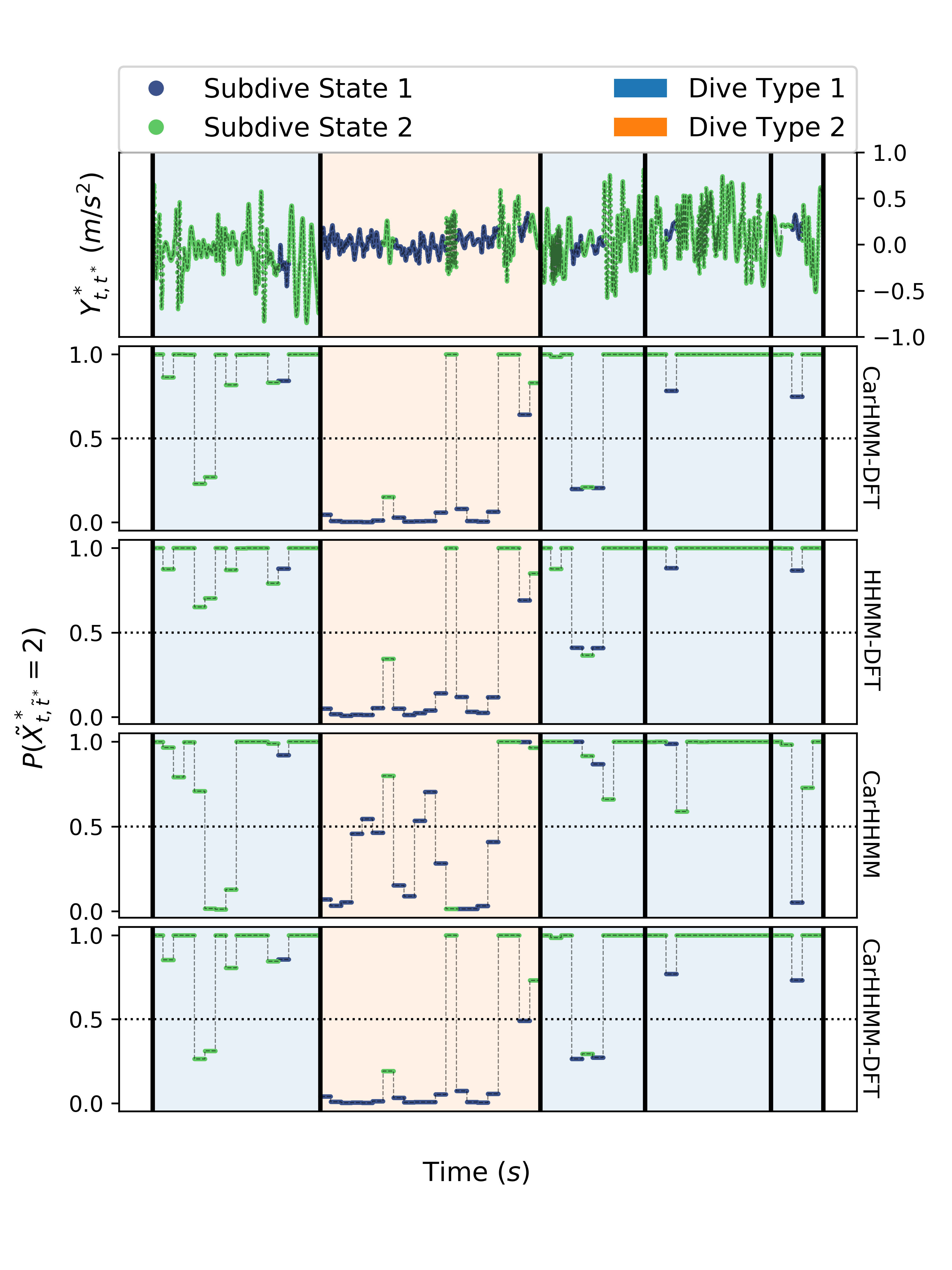}
    \caption{Estimated probabilities that each window corresponds to subdive state 2 for five selected dives of a simulated data set of killer whale dive behaviour. Each panel is partitioned into dives by vertical black lines. The colour of the curve corresponds to the true subdive state while the colour of the background corresponds to the true dive type.}
    \label{fig:acc_fine}
\end{figure}

The full CarHHMM-DFT is the best performing model of the four candidates since it is the generating model. Its average dive decoding accuracy and average subdive decoding accuracy are both greater than 0.9. All parameter estimates of fine-scale mean values and probability transition matrices, $\hat \mu^*$, $\hat \Gamma^*$ and $\hat \Gamma$, respectively, have statistically insignificant biases. In addition, the biases of $\hat \sigma$, $\hat \phi$, and $\hat \mu$ are either as small or smaller than the other three candidate models. The empirical standard errors of all parameter estimates ($\hat \theta$, $\hat \Gamma$, $\hat \theta^*$, $\hat \Gamma^*$) are well-approximated by the inverse of the observed Fisher information matrix, although the estimated standard errors tend to be slightly smaller than the empirical standard errors. This underestimation is especially noticeable for parameters associated with the wiggliness $\Ztwo_{t,\tilde t^*}$, where the empirical standard error can be up to twice as large as the estimated standard error. See Tables 5 through 9 of the supplementary material for detailed results.

The HHMM-DFT performs similarly to the CarHHMM-DFT in most respects. Its average dive decoding accuracy is comparable to the CarHHMM-DFT while its average subdive decoding accuracy is worse by approximately 5 percentage points. Its parameter estimates are comparable to the CarHHMM-DFT with the notable exception that it greatly overestimates $\sigma_A^{*(\cdot,1)}$ and $\sigma_A^{*(\cdot,2)}$. In addition, the estimated standard errors of $\hat \mu_A^{*(\cdot,1)}$, $\hat \mu_A^{*(\cdot,2)}$, $\hat \sigma_A^{*(\cdot,1)}$, and $\hat \sigma_A^{*(\cdot,2)}$ are much smaller than the associated empirical standard errors (see Table 7 of the supplementary material). These results suggest that the estimates of standard deviation can be too large and estimates of standard errors can be too small when auto-correlation is ignored. This finding is consistent with the results of the case study, where the HHMM-DFT produced larger estimates of $\sigma_A^*$ and smaller estimates of standard error compared to the CarHHMM-DFT. When standard errors are underestimated, the associated confidence intervals are too narrow, implying that researchers may be overconfident in their parameter estimates.

The CarHHMM is the worst-performing model in terms of accuracy, as its average dive decoding accuracy is below $0.9$ and its average subdive decoding accuracy is below $0.8$. This result is consistent with expectations because the CarHHMM does not model the ``wiggliness" of the fine-scale process, which is the most distinct difference between the subdive states. In addition to its poor average decoding accuracy, the CarHHMM is also the worst of the four candidate models at estimating parameters. Parameter estimates associated with subdive type 2 ($\theta^{*(\cdot,2)}$) are especially poor. See Section 2 of the supplementary material for more detailed results.

Finally, the CarHMM-DFT is nearly identical to the CarHHMM-DFT in terms of average subdive decoding accuracy, fine-scale parameter biases, and both estimated and empirical standard error for the fine-scale parameter estimates. In addition, the time required to fit the CarHMM-DFT is less than half of that of the other models (see Table \ref{table:accuracy}). However, this model cannot estimate dive type as it lacks any hierarchical structure. The CarHMM-DFT nonetheless fits a (misspecified) single Gamma distribution over the dive duration of all dives. The resulting parameter estimates ($\hat \mu$ and $\hat \sigma$) are highly correlated (See Figure 25 of the supplementary material).

Figures \ref{fig:acc_coarse} and \ref{fig:acc_fine} display five dives of one simulated data set as well as the decoded dive types and subdive states associated with each model. The CarHHMM-DFT and CarHMM-DFT produce similar estimates of subdive state while the HHMM-DFT is slightly more likely to predict that a given window corresponds to subdive state 2. As expected, the CarHHMM is the least accurate model when predicting subdive state. The CarHHMM-DFT and HHMM-DFT yield similar estimates of dive type while the CarHHMM is less accurate and even misclassifies the dive type of the second dive. The CarHMM-DFT does not estimate dive type.
\section{DISCUSSION}
\pdfoutput=1


Current Functional Data Analysis literature addresses dependence between curves either with multilevel models \citep{Chen:2012,Di:2009}, which lack a time component, or with functional time series, which overlook the possibility that curves have several distinct ``types" \citep{Kokoszka:2018}. Our work addresses these issues and introduces a flexible framework to model functional time-series data using HMMs.
We suggest handling temporal dependence between curves by using either an HMM or a CarHMM to model the curve sequence. We then suggest viewing each individual curve as an HMM emission whose distribution is described by a fine-scale model. Here we use a CarHMM as the fine-scale model, but there are a wide range of possible fine-scale models, including a Poisson process or continuous time approach similar to that of \citet{Michelot:2019}. We also incorporate a moving-window transformation at the fine scale to capture intricate dependence structures on short time scales. Together, the coarse- and fine-scale models make up a hierarchical structure which can account for simultaneous processes taking place at different time scales. Provided the construction is not overly complex, a hierarchical model created using our method can be both flexible and easy to fit using maximum likelihood estimation.

We demonstrate the usefulness of this framework using a biomechanical/ecological example, where we use HMMs to classify the coarse- and fine-scale diving behaviour of a northern resident killer whale in Queen Charlotte Sound off the coast of British Columbia, Canada. Our analysis gives a deeper understanding of a killer whale's tri-axial movement and thus its behaviour and energy expenditure \citep{Gleiss:2011,Qasem:2012}, both of which are important for understanding the foraging ecology and nutritional status of northern resident killer whales \citep{Noren:2011}. Our model is also applicable to many diving animals such as sharks \citep{Adam:2019}, seals \citep{Dot:2016}, and porpoises \citep{Barajas:2017}. In addition, since complicated state-switching processes with temporal dependence are common in settings ranging from speech recognition \citep{Juang:1991} and neuroscience \citep{Langrock:2013} to oceanography \citep{Bulla:2012} and ecology \citep{Adam:2019}, we believe that researchers can adapt our methodology for the analysis of a wide range of time series data in a variety of fields.

\section{ACKNOWLEDGEMENTS}
The killer whale data was collected under University of British Columbia Animal Care Permit no. A19-0053 and Fisheries and Oceans Canada Marine Mammal Scientific License for Whale Research no. XMMS 6 2019.
This research was enabled in part by support provided by WestGrid (www.westgrid.ca) and Compute Canada (www.computecanada.ca).
We acknowledge the support of the Natural Sciences and Engineering Research Council of Canada (NSERC) as well as the support of Fisheries and Oceans Canada (DFO). 
Marie Auger-M\'eth\'e and Nancy Heckman thank the NSERC Discovery program, and Marie Auger-M\'eth\'e additionally thanks the Canadian Research Chair program.
Evan Sidrow thanks the University of British Columbia and the Four-Year Doctoral Fellowship program.
We are grateful to Dr. Joe Watson for his constructive suggestions.
\newpage
\pdfoutput=1

%
\newpage
\begin{appendix}
\pdfoutput=1

\section{Detailed description of data simulation from Section \ref{subsec:data_simulation}}

\setcounter{equation}{4}   


We easily simulate realizations of the coarse-scale HMM ($X$ and $Y$) given the parameters $\Gamma$ and $\theta$. For each dive $t$, we also easily generate the fine-scale hidden Markov chain $\tilde X^*_t \equiv \left\{\tilde X^*_{t,1},\ldots,\tilde X^*_{t,\tilde T_t^*}\right\}$ according to one of the probability transition matrices $\Gamma^{*(1)}$ or $\Gamma^{*(2)}$, depending upon the value of $X_t$. This determines the sequence of fine-scale hidden states corresponding to each window. Recall that the fine-scale model is based on a sequence of $\tilde T_t^*$ two-second windows, each containing 100 observations, and our model is formulated in terms of quantities derived from the raw data within each window (namely, the average acceleration, $\Zone_{t,\tilde t^*}$, and wiggliness, $\Ztwo_{ t,\tilde t^*}$). Generating the raw acceleration data from $\Zone_{t,\tilde t^*}$ and $\Ztwo_{ t,\tilde t^*}$ is not straightforward. In our simulation study, we generate raw acceleration data so that we can visualize our results in terms of the underlying data curves for each dive. Here, we explain how we generate the acceleration curves so that $\Zone_{t,\tilde t^*}$ and $\Ztwo_{ t,\tilde t^*}$ both follow the specified model. A key component is the discrete Fourier transformation of the 100 raw acceleration values in window $\tilde{t}^*$ of dive $t$:
\[
    \hat{Y}^{*(k)}_{t,\tilde{t}^*}  \equiv DFT\left\{Y^*_{t,100 (\tilde{t}^*-1) + 1 },\ldots, Y^*_{t,100 \tilde{t}^*}\right\}(k)
\]
for $k \geq 0$, as defined in Equation (\ref{eq:DFTdef}).

We simulate the raw acceleration data for dive $t$ in three steps: (1) simulate the average acceleration within each window $\left(\hat Y^{*(0)}_{t,\tilde t^*}\right)$, (2) simulate all other Fourier coefficients within each window $\left(\hat Y^{*(k)}_{t,\tilde t^*}, k = 1,\ldots,99\right)$, and (3) take the inverse discrete Fourier transform of $\hat{Y}^*_{t,\tilde t^*}$:
\[
    \{Y^*_{t,100(\tilde t^* - 1) + 1},\ldots,Y^*_{t,100\tilde t^*}\} \equiv IDFT\left\{\hat{Y}^{*(0)}_{t,\tilde t^*},
    \ldots, \hat{Y}^{*(99)}_{t,\tilde t^*}\right\}
    {\rm{~for~}} \tilde t^* = 1,\ldots,\tilde T^*_t.
\]
The details of steps (1) and (2) are described below.

For step (1), we generate $\hat Y^{*(0)}_{t,1}, \ldots, \hat Y^{*(0)}_{t,\tilde T_t^*}$ as a CarHMM with underlying Markov state sequence $\tilde X^{*}_{t,1}, \ldots, \tilde X^{*}_{t,\tilde T_t^*}$ and random first emission. Specifically, we let
   \[
    	\hat{Y}^{*(0)}_{t,1}|\tilde X^*_{t,1} = i^* ~~\sim~~ \mathcal{N} \left(0, \left(100\sigma_A^{*(\cdot,i^*)}\right)^2 \right)
    	~~{\rm{and}}
   \]
     \begin{align}	
       \hat{Y}^{*(0)}_{t,\tilde t^*}|\tilde X^*_{t,\tilde t^*} = i^*,\hat{Y}^{*(0)}_{t,\tilde t^*-1}
       &~~\sim ~~\mathcal{N} \left(\phi_A^{*(\cdot,i^*)} \hat{Y}^{*(0)}_{t,\tilde t^*-1}, \left(100\sigma_A^{*(\cdot,i^*)}\right)^2 \right), \label{eqn:yhat_0} \\
        &\tilde t^* = 2,\ldots, \tilde T^*_t
    	\nonumber
    \end{align}
    where $\sigma_A^{*(\cdot,1)} = 0.034s$,  $\sigma_A^{*(\cdot,2)} = 0.079s$. $\phi_A^{*(\cdot,1)} = 0.98$ and $\phi_A^{*(\cdot,2)} = 0.87$. 
  
 For step (2),
 we first construct $\hat{Y}^{*(k)}_{t,\tilde t^*}$, $k=1,\ldots, 49$, as
    \begin{equation}
        \hat{Y}^{*(k)}_{t,\tilde t^*} = a_{t,\tilde t^*}^{(k)} i\sqrt{b^{(k)}_{t,\tilde t^*}}
        \label{eqn:abYhat}
    \end{equation}
    where the $a^{(k)}_{t,\tilde t^*}$'s are independent and equal to either 1 or -1 each with probability 1/2 and $i$ is the imaginary number. We include $i$ in Equation (\ref{eqn:abYhat}) to force all variation within a window to take the form of a sine wave, which reduces the variation between the endpoints of windows compared to a cosine wave.
    Given the fine scale states, the $b^{(k)}_{t,\tilde t^*}$'s are independent and independent of the  $a^{(k)}_{t,\tilde t^*}$'s.  
   The distribution of $b^{(k)}_{t,\tilde t^*}$ is
    \begin{align}
    \begin{split}
    	b^{(k)}_{t,\tilde t^*}|\tilde X^*_{t,\tilde t^*} = 1 &~~\sim~~ {\rm{Gamma}}(16.38/k^3, 36.23) \\
    	b^{(k)}_{t,\tilde t^*}|\tilde X^*_{t,\tilde t^*} = 2  &~~\sim ~~{\rm{Gamma}}(4.20/k^3, 1825.53). \\ 
    \end{split}
    \label{eqn:bdist}
    \end{align}
    The first argument of ${\rm{Gamma}}\left(\cdot,\cdot\right)$ is the shape parameter and the second is the scale parameter. The squared magnitude of the $k^{th}$ Fourier coefficient is equal to $b^{(k)}_{t,\tilde t^*}$, which decays like $1/k^3$ to ``smooth out" the raw acceleration data.
    
   We then define the remaining 50 Fourier coefficients:
    \[
    \hat{Y}^{*(50)}_{t,\tilde t^*} = 0 {\rm{~~and ~~}}
	\hat{Y}^{*(k)}_{t,\tilde t^*} = -\hat{Y}^{*(100-k)}_{t,\tilde t^*} , k = 51,\ldots,99.
\]
   This guarantees that the inverse discrete Fourier transform is real-valued.

We now show that this construction of the raw acceleration data results in the distributions listed in Section \ref{subsec:data_simulation}. It suffices to show that the construction of the discrete Fourier transformations, the $\hat{Y}^{(k)}_{t,\tilde t^*}$'s, yields the desired distributions.

First, since $\hat{Y}^{(0)}_{t,\tilde t^*} = \sum_{n=1}^{100} Y^*_{t,100(\tilde t^* - 1) + n} $ $=100\Zone_{t,\tilde{t}^*}$, Equation (\ref{eqn:yhat_0}) implies that 
$\Zone_{t,1}, \ldots, \Zone_{t,\tilde{T}^*}$ 
follows a CarHMM with Normal emissions distributions and parameters 
 $\mu_A^{*(\cdot,1)} = \mu_A^{*(\cdot,2)} = \mu_A^{*(\cdot,i^*)} = 0$,  $\sigma_A^{*(\cdot,1)} = 0.034s$, $\phi_A^{*(\cdot,1)} = 0.98$, $\sigma_A^{*(\cdot,2)} = 0.079s$, and $\phi_A^{*(\cdot,2)} = 0.87$.

 From Equations (\ref{eqn:z}) and (\ref{eqn:abYhat}), 
 the wiggliness within window $\tilde t^*$ of dive $t$ is 
\[
    \Ztwo_{t,\tilde t^*} = \sum_{k=1}^{\tilde \omega} \big|\big|\hat{Y}^{(k)}_{t,\tilde t^*}\big|\big|^2 =  \sum_{k=1}^{\tilde \omega} b^{(k)}_{t,\tilde t^*}.
\]
If $\tilde \omega < 50$, then $\Ztwo_{t,\tilde t^*}$ is the sum of independent Gamma-distributed random variables with identical scale parameters, so the distribution of $\Ztwo_{t,\tilde t^*}$ is also Gamma. Thus, by Equation (\ref{eqn:bdist})
\[
\Ztwo_{t,\tilde t^*}|\tilde X^*_{t,\tilde t^*} = 1  ~~ \sim ~~ {\rm{Gamma}}\left(\sum_{k=1}^{\tilde \omega} 16.38/k^3 , 36.23 \right) 
\text { and }
\]
\[
\Ztwo_{t,\tilde t^*}|\tilde X^*_{t,\tilde t^*} = 2 ~~\sim~~ {\rm{Gamma}}\left(\sum_{k=1}^{\tilde \omega} 4.20/k^3 , 1825.53 \right).
\]
Setting $\tilde \omega$ to 10 and carrying out simple calculation of the mean and variance of a Gamma distribution yields $\mu_W^{*(\cdot,1)} = 23.3$, $\sigma_W^{*(\cdot,1)} = 13.0$, $\mu_W^{*(\cdot,2)} = 301.2$, and $\sigma_W^{*(\cdot,2)} = 330.1$.


\section{Likelihood of CarHHMM-DFT}

The overall likelihood of the CarHHMM-DFT model is as follows:
$$\calL_{\text{CarHHMM-DFT}}(\theta,\theta^*,\Gamma,\Gamma^*;y,\z) = \delta P(y_1,\z_1;\theta,\theta^*,\Gamma^*) \prod_{t=2}^T \Gamma P(y_t,\z_t;\theta,\theta^*,\Gamma^*) \mathbf{1}_N.$$
In particular,
\begin{align*}
P(y_t,\z_t;\theta,\theta^*,\Gamma^*)  = \text{diag}\Big[&f^{(1)}(y_t;\theta^{(1)})\calL_{\text{fine}}\left(\theta^*,\Gamma^{*(1)};\z_t\right), \ldots , \\
&f^{(N)}(y_t;\theta^{(N)})\calL_{\text{fine}}\left(\theta^*,\Gamma^{*(N)};\z_t\right) \Big],
\end{align*}
where $f^{(i)}(y_t;\theta^{(i)})$ is the emission distribution of dive duration given that $X_t = i$. The likelihood $\calL_{\text{fine}}$ corresponds to the fine-scale model and is equal to the following:
$$\calL_{\text{fine}}\left(\theta^{*},\Gamma^{*(i)};\z_t\right) = \delta^{*(i)} \prod_{\tilde t^* = 2}^{\tilde T^*_t} \Gamma^{*(i)} P(\z_{t,\tilde t^*}|\z_{t,\tilde t^*-1};\theta^*) \mathbf{1}_{N^*},$$
where $P(\z_{t,\tilde t^*}|\z_{t,\tilde t^*-1};\theta^*)$ is an $N^* \times N^*$ diagonal matrix with $(i^*,i^*)^{th}$ entry equal to $f^{*(\cdot,i^*)}(\z_{t,\tilde t^*}|\z_{t,\tilde t^*-1}; \theta^{*(\cdot,i^*)})$.
Recall that $f^{*(\cdot,i^*)}(\cdot|\z_{t,t^*-1}; \theta^{*(\cdot,i^*)})$ is the probability density function of $\Z_{t,\tilde t^*}$ when $\tilde X^*_{t,\tilde t^*} = i^*$ and $\Z_{t,\tilde t^*-1} = \z_{t,\tilde t^*-1}$.


\end{appendix}

\includepdf[pages=1-]{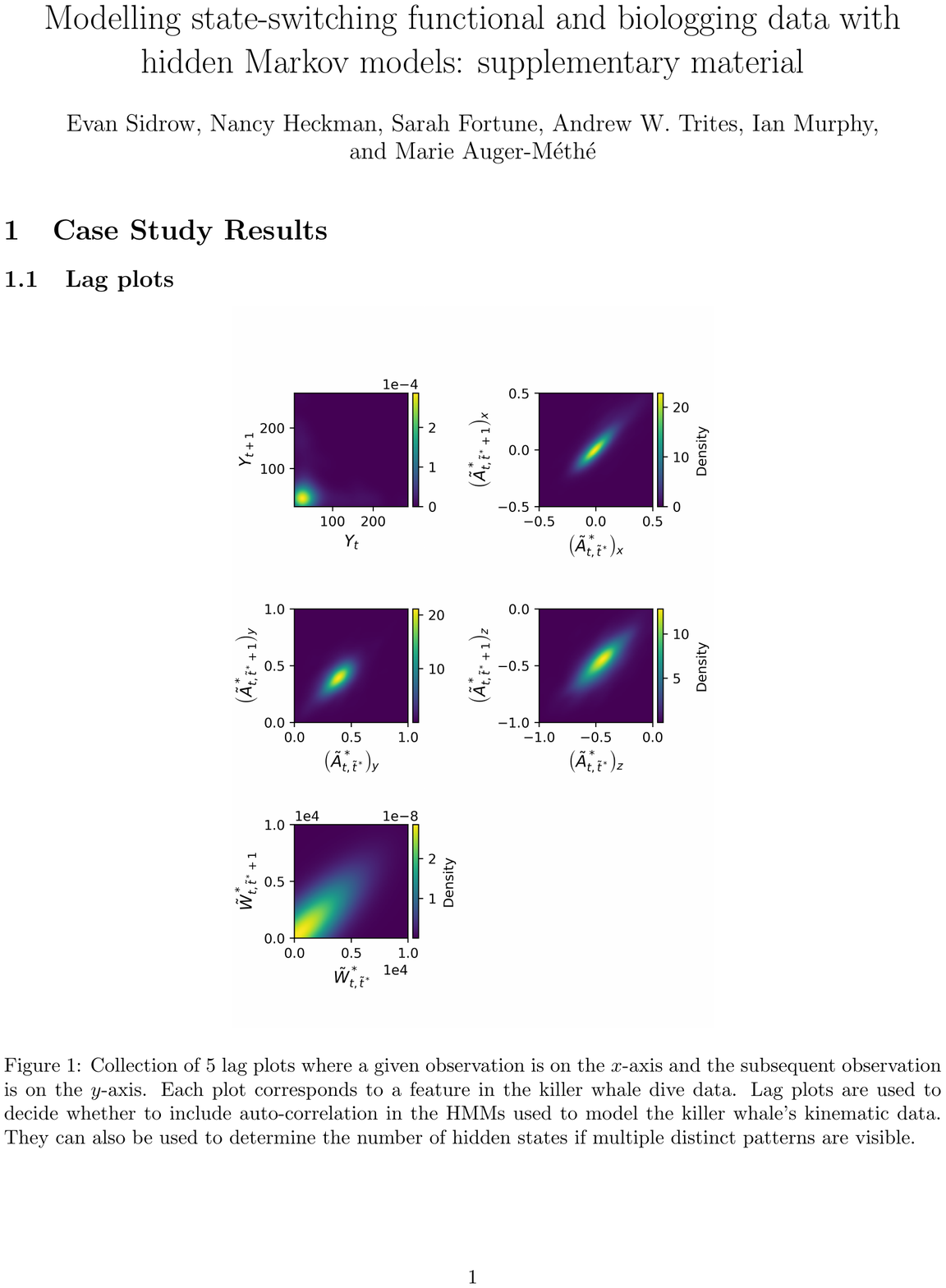}

\end{document}